\documentclass[10pt]{article}
\pdfoutput=1
\usepackage{jcappub}

\usepackage[usenames]{xcolor}
\usepackage[english]{babel}
\usepackage[small]{caption}
\usepackage{verbatim}
\usepackage{multirow}
\usepackage{relsize}
\usepackage{amsmath,amssymb,amsbsy,amstext,amsfonts}
\usepackage{epsfig,multicol,bbm,simplewick}
\usepackage{mathtools}
\usepackage{physics}
\usepackage{enumitem}
\usepackage{csquotes}
\usepackage{tensor}
\usepackage{fix-cm}
\usepackage{booktabs}
\usepackage{graphicx}
\usepackage{bm}
\usepackage{subfigure}
\usepackage{float}
\usepackage[explicit]{titlesec}
\usepackage{enumitem}
\usepackage{setspace}
\usepackage{blindtext}
\usepackage{textpos}
\usepackage{etoolbox}
\usepackage[linktocpage=true]{hyperref}
\usepackage{anyfontsize}
\usepackage{listings,xfp}
\usepackage[dotinlabels]{titletoc}
\usepackage[normalem]{ulem}
\usepackage{esvect}
\DeclareMathAlphabet\mathbfcal{OMS}{cmsy}{b}{n}
\usepackage{tikz}
\usepackage{tikz-feynman}
\usepackage{soul}
\usepackage{subcaption}
\usepackage{orcidlink}

\definecolor{jrcolor}{rgb}{.2,.0,.8}

\definecolor{airforceblue}{rgb}{0.36, 0.48, 0.84}
\definecolor{purplemath}{rgb}{0.5, 0, 0.5}

\hypersetup{colorlinks=true,linkcolor=blue,citecolor=blue,urlcolor=blue}
\usepackage[a4paper, total={6.75in, 9.75in}]{geometry}
\renewcommand{\thesection}{\arabic{section}}

\makeatletter
\def\@xfootnote[#1]{%
  \protected@xdef\@thefnmark{#1}%
  \@footnotemark\@footnotetext}
\makeatother

\def\bp{\boldsymbol{p}}
\def\bP{\boldsymbol{P}}
\def\bk{\boldsymbol{k}}
\def\bK{\boldsymbol{K}}
\def\bx{\boldsymbol{x}}

\def\ba{\boldsymbol{a}}
\def\bb{\boldsymbol{b}}

\newcommand{\beq}{\begin{equation}\begin{aligned}}
\newcommand{\eeq}{\end{aligned}\end{equation}}

\newcommand{\bea}{\begin{eqnarray}}
\newcommand{\eea}{\end{eqnarray}}

\newcommand*\diff{\mathop{}\!\mathrm{d}}
\newcommand{\overbar}[1]{\mkern 1.5mu\overline{\mkern-1.5mu#1\mkern-1.5mu}\mkern 1.5mu}

\title{Fermion (non)reheating with a quartic inflaton potential}

\author[a]{Nabeen Bhusal \orcidlink{0009-0006-8262-3691},}
\affiliation[a  ]{Deutsches Elektronen-Synchrotron DESY,\\ Notkestr.~85, 22607 Hamburg, Germany} 

\author[b,c]{Ernesto Ch\'avez M. \orcidlink{0009-0007-8857-1463 },}
\affiliation[b  ]{Departamento de F\'isica Te\'orica, Instituto de F\'isica, \\
Universidad Nacional Aut\'onoma de M\'exico, \\
Ciudad de M\'exico C.P. 04510, Mexico}
\affiliation[c  ]{Instituto de Astronom\'ia, \\
Universidad Nacional Aut\'onoma de M\'exico, \\
Ciudad de M\'exico C.P. 04510, Mexico}

\author[b]{Marcos A. G. Garcia \orcidlink{0000-0003-3496-3027},}

\author[a]{Adriana G. Menkara \orcidlink{0000-0001-7974-1909},}

\author[a]{and Mathias Pierre \orcidlink{0000-0002-2159-1845}}

\emailAdd{nabeen.bhusal@desy.de}
\emailAdd{iechavez@astro.unam.mx}
\emailAdd{marcos.garcia@fisica.unam.mx}
\emailAdd{adriana.menkara@desy.de}
\emailAdd{mathias.pierre@desy.de}

\abstract{Any viable inflationary model must account for reheating of the universe prior to the onset of primordial nucleosynthesis. In this work, we study the (p)reheating mechanism for an inflaton field with a quartic minimum of the T-model kind with coupling $\lambda$, prior to and post fragmentation, making a clear distinction between the two regimes. We assume that the main particle production channel corresponds to the decay into a pair of spin 1/2 fermions via Yukawa-like interactions. On top of its decays, we also consider the self-interaction of the inflaton, which sources the resonant growth of inflaton inhomogeneities, possibly leading to its eventual fragmentation. By means of a combination of non-perturbative (Heisenberg/Bogoliubov) and perturbative (Boltzmann) methods, we find that for Yukawa couplings that seemed to be intuitively perturbative, such as $y\gtrsim 10^{-8}$ ($y^2/\lambda\gtrsim 3\times10^{-5}$), parametric resonance, kinematic blocking, and Pauli suppression effects cannot be ignored. Additionally, we show that achieving  $\rho_\phi \sim \rho_\psi$ prior to fragmentation requires large couplings, $y\gtrsim 0.2$ ($y^2/\lambda\gtrsim 10^{10}$), which needs a detailed study of backreaction and radiative corrections. Thus the rest of our work constitutes studying post-fragmentation fermion production where we conclude that, in general, reheating in this setup is not possible and thus we conclude that in order to successfully reheat, one must invoke a coupling to a integer- and/or 0-spin particle like a scalar boson.}

\begin{document}
\begin{flushright}
	\footnotesize
	DESY-25-192 \\
\end{flushright}
\color{black}
\maketitle
\flushbottom


\section{Introduction}

The proposed early epoch of accelerated expansion, known as cosmic inflation, is a powerful paradigm for solving the initial conditions problems of the standard Big Bang scenario. Besides explaining the homogeneity, isotropy, and flatness of the universe, it provides a causal mechanism for seeding the initial density fluctuations necessary to explain the Cosmic Microwave Background (CMB) temperature and polarization anisotropies, and the observed large scale structure of the universe~\cite{Olive:1989nu,Linde:1990flp,Lyth:1998xn,Linde:2000kn}. In the slow-roll formalism, the source of the accelerated expansion and the initial inhomogeneities is a real neutral scalar field $\phi$, known as the {\em inflaton}, and its quantum fluctuations with respect to a time-dependent homogeneous background value, respectively. Plateau-like potentials, such as the original Starobinsky model~\cite{Starobinsky:1980te} or the more recently proposed T-models with potential~\cite{Kallosh:2013hoa}
\beq
V(\phi) \; = \, \lambda  M_P^4 \left(\sqrt{6} \tanh \left(\frac{\phi }{\sqrt{6}M_P}\right)\right)^k,\;
\eeq
predict scalar and tensor spectra compatible with the observational data retrieved by the {\em Planck} and BICEP/Keck observatories~\cite{Planck:2018jri,BICEP:2021xfz,Tristram:2021tvh}, although in relative tension with the more recent Atacama Cosmology Telescope (ACT) and South Pole Telescope (SPT) results~\cite{ACT:2025fju,ACT:2025tim,SPT-3G:2025bzu,Ellis:2025zrf}. 

The energy density of the inflaton and its inhomogeneities are transferred to the primordial relativistic plasma after the end of slow-roll during the reheating epoch. Under the assumption of small inflaton-matter and inflaton-inflaton couplings, the expansion of the universe during reheating is determined by the underdamped oscillation of the inflaton field about the minimum of its potential. For the T-models, with $V\sim \phi^k$ for $\phi\ll M_P$, these oscillations mimic a matter-dominated universe for $k=2$, a radiation-dominated universe for $k=4$, and possess a stiff equation of state for $k>4$~\cite{Turner:1983he,Martin:2010kz,Garcia:2020wiy}. Under these approximations the energy density stored in these oscillations is slowly transferred into rapidly thermalising relativistic particles until its depletion, signalling the end of reheating and the beginning of the radiation dominated era. Assuming that the primary particle production process corresponds to inflaton decays, the transfer rate of energy can be directly associated to the decay rate of the inflaton particle. 

Generically, however, the coherence of the oscillations of the inflaton field is not maintained until the end of reheating. Even considering weak inflaton-matter couplings, for $k>2$ the self-interactions of the inflaton are sufficiently effective to drive the growth of the inflaton inhomogeneities via the phenomenon of parametric resonance~\cite{Dolgov:1989us,Traschen:1990sw, Shtanov:1994ce, Boyanovsky:1995ud, Yoshimura:1995gc, Kofman:1997yn}. The eventual result is the fragmentation of the inflaton condensate in favour of an incoherent collection of inflaton quanta, with non-thermal Phase Space Distributions (PSD) and effective masses that continuously decrease with time. Reheating after fragmentation requires then the decay of these inflaton particles into light degrees of freedom. Depending on the spin of the daughter particles, these decays may be roughly insensitive to the fragmentation of $\phi$ (e.g.~scalars) or may be strongly suppressed compared to the direct production from the condensate (e.g.~spin 1/2 fermions)~\cite{Garcia:2023eol,Garcia:2023dyf,Garcia:2024rwg}. In general, including the case for $k=2$, the fragmentation of the inflaton field can also be originated from the resonant growth of bosonic fields strongly coupled to the inflaton~\cite{Greene:1997fu,Kaiser:1997mp,Felder:2006cc, Garcia-Bellido:2008ycs, Frolov:2010sz,Amin:2011hj,Hertzberg:2014iza,Amin:2014eta,Lozanov:2016hid,Lozanov:2017hjm, Figueroa:2016wxr,Fu:2017ero,Antusch:2021aiw,Lebedev:2022vwf,Garcia:2021iag}.

In the present work, we focus on the production of spin $1/2$ fermions from the decay of the inflaton field as it oscillates about a quartic minimum, choosing for definiteness the quartic T-model.  We provide for the first time a simultaneous treatment of fermion production and inflaton fragmentation. Earlier studies of fermionic preheating assume a spatially homogeneous oscillating inflaton background~\cite{Giudice:1999fb,Greene:1998nh,Kuzmin:1998kk,Garcia-Bellido:2000woy,Peloso:2000hy,Nilles:2001fg,Chung:2011ck,Adshead:2018oaa,Boyle:2018rgh,Garcia:2021iag,Casagrande:2023fjk,Kolb:2023ydq}. Conversely, studies of inflaton fragmentation usually focus on bosonic decay channels, or very superficially treat fermions. Our work aims to provide a consistent study of fermionic production from inflation that accounts for the production from the condensate as well as from fragmentation.

A prerequisite for reheating is the equality between the energy densities of the inflaton and its daughter products. In this work, we investigate whether this minimal requirement can be met in the first place.
This allows us to set lower bounds on the strength of the Yukawa-like coupling. We take on the task of computing the fermion production from the respective dominant mechanism prior to and post-fragmentation in an attempt to provide a more complete picture of fermion production during the preheating stage after inflation. Our results suggest that in fact, fermions are unlikely to reheat and thus, in quartic inflaton potentials, a scalar or a vector boson must be additionally coupled to fermions. Unlike in a previous attempt ~\cite{Garcia:2023eol}, we study here the dynamics of the fermionic sector without making the perturbative (Boltzmann) approximation, requiring for simplicity only the presence of the inflaton and the fermion sectors, and a direct coupling between them. This means that we account for the decay of the condensate inflaton component and the inhomogeneous inflaton fluctuations before and after fragmentation. It also means that, prior to fragmentation, we follow the fermion field by numerically solving its Dirac equation in the weak and strong coupling regimes, obtaining its PSD and the corresponding energy density. After fragmentation, in order to account for free inflaton quanta decays, we take the previously obtained fermion PSD as the initial condition for its further evolution using the Boltzmann equation. We solve this equation numerically to ensure that the effects of time-dependent masses and fermionic quantum statistics are properly taken into account. We demonstrate that, in the range of couplings identified in~\cite{Garcia:2023eol} as required for successful reheating, the presence of resonance effects, kinematic blocking and Pauli suppression can strongly halt the decays of the inflaton, further reducing the parameter space for successful fermionic reheating. We find in particular that in the absence of other inflaton-matter couplings, reheating in a quartic potential cannot be fully completed by its perturbative decay into fermions.

The structure of the paper is the following. In Section~\ref{sec:afterinflation}, we introduce the inflation model on which we base our study, and the couplings between the inflaton and the spin 1/2 fermion that will be responsible for reheating. Section~\ref{sec:infosc} summarizes the background evolution of the homogeneous inflaton condensate prior to fragmentation, and the equations that govern the dynamics of the fermion sector. Section~\ref{sec:frag} provides a brief summary of inflaton fragmentation in a quartic potential. Our analysis of fermion production during reheating in a quartic potential prior to fragmentation is presented in Section~\ref{eq:beforefrag}. In Section~\ref{sec:boltzmann}, we make use of the Boltzmann equation to construct a perturbative approximation to the dynamics, which is contrasted with the full non-perturbative (Heisenberg) solution in Section~\ref{sec:nonpert}. This section contains our exploration of the parameter space for successful reheating prior to inflaton fragmentation. Section~\ref{sec:afterfrag} presents our analysis of particle production after inflaton fragmentation without any approximations, other than the use of the Boltzmann equation for the decay of free inflaton quanta. Finally, our conclusions and a summary of our findings are presented in Section~\ref{sec:conclusion}.

\section{Dynamics after inflation}\label{sec:afterinflation}

The field content for our study consists on the inflaton field $\phi$ and a spin 1/2 fermion $\psi$ which is part of, or communicates with, the visible sector of the universe. The corresponding action is written as
\beq
\label{eq:action}
\mathcal{S} \;=\; \int \diff ^4x\,\sqrt{-g} \left[ -\dfrac{M_P^2}{2} R + \frac{1}{2}(\partial_{\mu}\phi)^2 - V(\phi) + \bar{\psi}(i\bar{\gamma}^{\mu}\nabla_{\mu} - m_{\psi} )\psi + \mathcal{L}_{\rm int} \right] \, ,
\eeq
where in the gravitational sector $g$ denotes the determinant of the metric tensor, $R$ is the Ricci scalar and $M_P=1/\sqrt{8\pi\,G}\simeq 2.45\times 10^{18}\,{\rm GeV}$ is the reduced Planck mass. For the inflaton sector we choose the quartic T-model potential
\begin{align}\label{eq:Tquart}
V(\phi) \; &= \, \lambda  M_P^4 \left(\sqrt{6} \tanh \left(\frac{\phi }{\sqrt{6}M_P}\right)\right)^4\\ \label{eq:Tquart2}
&\simeq\; \lambda\phi^4\quad\ \mathrm{for}\ \ \phi\ll M_P.
\end{align}
The value of the parameter $\lambda$ is fixed by the measurement of the amplitude of the primordial curvature spectrum at the {\em Planck} pivot scale $k_*=0.05\,{\rm Mpc}^{-1}$, $\ln(10^{10}A_{S*})=3.044$~\cite{Planck:2018jri,Planck:2018vyg}. In the slow-roll approximation, 
\beq\label{eq:lambda}
\lambda\;\simeq\; \frac{\pi^2A_{S*}}{2N_*^2}\,,
\eeq
where the number of $e$-folds between horizon exit of the pivot scale and the end of inflation can be related to the corresponding field values as
\begin{equation}
    N_* \;\simeq\; \frac{1}{M_P^2}\int_{\phi_{\rm end}}^{\phi_*} \frac{V(\phi)}{V'(\phi)}\,\diff \phi \;=\; \dfrac{3}{8} \left[ \cosh \left( \sqrt{\dfrac{2}{3}} \phi_* \right) - \cosh \left( \sqrt{\dfrac{2}{3}} \phi_{\rm end} \right) \right]\,,
    \label{eq:phistar}
\end{equation}
and with $\phi_*$ being the inflaton field value at horizon crossing of $k_*$, and $\phi_{\rm end}$ the field value at the end of inflation. For the {\em Planck} pivot scale, $N_*\simeq 56$ and $\lambda \simeq 3.3\times 10^{-12}$~\cite{Garcia:2020wiy,Ellis:2025zrf}.

For the fermion sector in (\ref{eq:action}) the curved-space gamma matrices $\bar{\gamma}^{\mu}$ are given in terms of the flat spacetime matrices as $\bar{\gamma}^{\mu}=e^{\mu}_a\gamma^a$, with $e^{\mu}_a$ the components of the metric tetrad.\footnote{We take the representation for the gamma matrices
\beq
\label{eq:notationgamma}
\gamma^0=  \begin{pmatrix}
    \mathbb{I},0\\
    0, \mathbb{I} 
\end{pmatrix} ~~~~,~~~~       
  \gamma^i =\begin{pmatrix}
    0 , \sigma_i\\
    -\sigma_i,0 
\end{pmatrix} ~~~~,~~~~       
\gamma^5 =\begin{pmatrix}
  0 , \mathbb{I}\\
  \mathbb{I},0 
\end{pmatrix} ~~
\eeq } .The covariant derivative for the spin 1/2 fields is defined as
\beq
\nabla_{\mu} \;=\; \partial_{\mu} - \frac{i}{2} \omega_{\mu ab}S^{ab}\,,
\eeq
where $\omega_{\mu ab}$ are the components of the spin connection, and the Lorentz algebra generators are given by $S^{ab}=\frac{i}{4}[\gamma^a,\gamma^b]$. $\mathcal{L}_{\rm int}$ represents interaction terms between the inflaton and $\psi$, which in this work we restrict to
\beq
-\mathcal{L}_\text{int} \, = \,  
y \phi \bar{\psi} \psi  + i y_5 \phi \bar \psi \gamma_5 \psi \,,
\label{couplingtofermions}
\eeq
where $y$ and $y_5$ are the scalar and pseudoscalar Yukawa couplings, respectively.

\subsection{Coherent inflaton oscillations}\label{sec:infosc}

The background dynamics during inflation and reheating are dominated by the inflaton field. Under the approximation that the inflaton inhomogeneities are negligible, well before the end of reheating, the classical inflaton condensate satisfies the coupled system of Klein Gordon-Friedmann equations for the field $\phi$ and its energy density $\rho_\phi$
\begin{align} \label{eq:KGF1}
\ddot{\phi}+3H\dot{\phi}+ \partial_{\phi} V   \;&=\;0\,,\\ \label{eq:KGF2}
\rho_{\phi} \;=\; \frac{1}{2}\dot{\phi}^2+V(\phi) \;&=\; 3H^2 M_P^2\,,
\end{align}
where $H= \dot{a}/a$ is the Hubble parameter, $a$ is the scale factor of the background Friedmann-Lema\^itre-Robertson-Walker (FLRW)
metric  and dots denote derivatives with respect to cosmic time. The transition between the inflationary accelerated expansion era and the stage of coherent oscillations during reheating is marked by the condition $\ddot{a}=0$, equivalent to the condition $\dot \phi_\text{end}^2=V(\phi_\text{end})$ at the end of inflation $a=a_\text{end}$. Solving for this condition over the inflationary attractor solution leads to
\beq
\phi_{\rm end} \;\simeq\; 1.52\,M_P\,,\quad \rho_{\rm end}\;\simeq\; (4.5\times 10^{15}\,{\rm GeV})^4\,,
\eeq
for the quartic T-model. After inflation, the solution to the system of equations (\ref{eq:KGF1})-(\ref{eq:KGF2}) can be approximated as the product of a decaying envelope function $\phi_0$ and a oscillating function $\mathcal{P}(t)$,
\beq
    \phi(t) \; = \; \phi_0 (t) \,\mathcal{P}(t).
\eeq
The time-dependence of the envelope function, and consequently the expansion rate of the universe, can be obtained averaging the dynamics over the underdamped oscillations. As it is well known, this averaging shows that the anharmonic oscillation of $\phi$ about the quartic potential mimics a radiation dominated universe, for which $\rho_{\phi}\propto a^{-4}$~\cite{Turner:1983he,Garcia:2020eof,Garcia:2020wiy}, and
\beq
 \phi_0 \; \simeq \; \phi_\text{end} \left( \dfrac{a_\text{end}}{a} \right)   \,.
\eeq
The effective mass of the inflaton is conventionally defined in terms of the envelope function as
\begin{equation}
    m_\phi^2(t) \, \equiv \, \partial_{\phi} ^2 V (\phi_0(t)) \, \simeq \, 12 \lambda  \phi_0^{2} \, \equiv \, m_{\rm end}^2\left( \dfrac{a_\text{end}}{a} \right)^{2} \,,
    \label{eq:mphi}
\end{equation}
which is a decreasing function in time. In terms of this effective mass, the oscillating function $\mathcal{P}(t)$ can be found in a closed form. Namely, introducing conformal time $ \diff \tau= \diff t/a$ we can write
\beq \label{eq:oscillating}
\mathcal{P}(\tau) \;=\; \mathrm{sn}\left( \frac{m_{\rm end}}{\sqrt{6}}(\tau-\tau_{\rm end}),-1) \right)\,,
\eeq
where sn denotes the Jacobi elliptic sine function~\cite{Garcia:2020wiy}.\par\medskip

To obtain the equation of motion for the fermion $\psi$ it is necessary to evaluate the metric tetrad and the spin connection in the FLRW background. In conformal time it is straightforward to show that $e_a^{\mu}=a^{-1}\delta_a^{\mu}$ and $-\frac{i}{2}\omega_{\mu ab}S^{ab}=\frac{a'}{4a^2}[\bar{\gamma}_{\mu},\gamma^0]$. Substitution into the equation of motion yields~\cite{Giudice:1999fb,Greene:1998nh,Kuzmin:1998kk,Garcia-Bellido:2000woy,Peloso:2000hy,Nilles:2001fg,Chung:2011ck,Adshead:2018oaa,Boyle:2018rgh,Garcia:2021iag,Casagrande:2023fjk,Kolb:2023ydq}
\beq\label{eq:Diraceq}
\left( i\gamma^{\mu}\partial_{\mu} + \frac{3ia'}{2a}\gamma^0 - a m_{\psi,{\rm eff}} \right)\psi \;=\; 0\,,
\eeq
where primes denote derivatives with respect to conformal time $\tau$, and 
\beq
m_{\psi,{\rm eff}} \;=\;  m_{\psi} + y \phi + i y_5 \phi  \gamma_5 \,,
\eeq
is the time-dependent effective mass matrix for the fermions. The equation of motion takes the more familiar form
\beq\label{eq:diracP}
\left( i\gamma^{\mu}\partial_{\mu} - a m_{\psi,{\rm eff}}\right)\Psi \;=\; 0\,,
\eeq
in terms of the canonically normalized field $\Psi=a^{3/2}\psi$. For it we introduce the mode expansion decomposed over spin states $s$
\beq
\Psi(\tau,\bx) \;=\; \sum_{s=\pm}\int \frac{\diff^3\bp}{(2\pi)^3} e^{-i\bp\cdot \bx}\,\left( u_p^{(s)}(\tau)\ba_{\bp}^{(s)} + v_p^{(s)}(\tau)\bb_{-\bp}^{(s)\dagger} \right)\,,
\eeq
where the ladder operators $\ba_{\bp}^{(s)},\bb_{\bp}^{(s)}$ satisfy the usual anticommutation relations, $\{\ba_{\bp}^{(s)},\ba_{\bp'}^{(s')\dagger}\}=\{\bb_{\bp}^{(s)},\bb_{\bp'}^{(s')\dagger}\}=\delta_{ss'}\delta^{(3)}(\bp-\bp')$ and all other anticommutators vanishing. The spinor mode functions are connected by the relation $v_p^{(s)}=\mathcal{C}\bar{u}_p^{(s)\,T}$, where $\mathcal{C}=i\gamma^2\gamma^0$ is the charge conjugation matrix. In the Dirac representation for the $\gamma$-matrices, denoting by $\boldsymbol{\sigma}$ the Pauli matrix vector, and by $\xi^s$ the eigenspinors of the helicity operator, $(\boldsymbol{\sigma}\cdot\hat{\bp})\xi^s=s\xi^s$, it is convenient to introduce the ansatz
\beq
u_p^{(s)}(\tau) \;=\; \left( \begin{matrix}
U_1^{(s)}(\tau) \xi^s(\bp)\\
U_2^{(s)}(\tau)(\boldsymbol{\sigma}\cdot\hat{\bp})\xi^s(\bp)
\end{matrix} \right)\,.
\eeq
Substitution into Eq.~(\ref{eq:diracP}) yields the equations of motion
\beq
\label{eq:diraceq1}
\begin{aligned}
U_1^{(s)\,\prime}(\tau) \;& =\; -i p U_2^{(s)}(\tau) - i a (m_\psi + y\phi) U_1^{(s)}(\tau) + s a y_5 \phi U_2^{(s)}(\tau)\,,\\
U_2^{(s)\, \prime}(\tau) \;& =\; -i p U_1^{(s)}(\tau) + i a (m_\psi + y\phi) U_2^{(s)}(\tau)  - s a y_5 \phi U_1^{(s)}(\tau)\,.
\end{aligned}
\eeq
These mode functions are subject to the constraint $|U_1^{(s)}|^2+|U_2^{(s)}|^2=1$ for consistency with the commutation relations of $\Psi$. The initial conditions are taken as the zero-particle Bunch-Davies mode functions~\cite{Adshead:2018oaa} for modes deep inside the horizon during inflation $p \ll a H$ 
\beq
U_1^{(s)}(\tau_0) \;=\; \frac{1}{\sqrt{2}}\sqrt{1-\frac{a(m_{\psi}+y\phi)}{\omega_p}}e^{is\theta}\,,\quad U_2^{(s)}(\tau_0) \;=\; \frac{1}{\sqrt{2}}\sqrt{1+\frac{a(m_{\psi}+y\phi)}{\omega_p}}\,,
\eeq
where
\beq
\omega_p\;\equiv\; \sqrt{p^2+a^2(m_{\psi}^2+y^2\phi^2+y_5^2\phi^2)}\qquad\text{and}\qquad e^{i\theta}\;\equiv\; \frac{p+iay_5\phi}{\sqrt{p^2+a^2y_5^2\phi^2}}\,.
\eeq

The energy density for the $\psi$ field is extracted from its energy-momentum tensor, which has the on-shell form~\cite{Casagrande:2023fjk} 
\beq
T^{\mu\nu} \;=\; \frac{i}{2}\bar{\psi}\bar{\gamma}^{(\mu}\overset{\leftrightarrow}{\nabla}{}^{\nu)}\psi\,,
\eeq
corresponding to
\beq
T^0_0 \;=\; \frac{i}{2a^4}\left( \bar{\Psi}\gamma^0\Psi'-\bar{\Psi}'\gamma^0\Psi\right)\,.
\eeq
Upon the introduction of Bogoliubov coefficients, and the computation of the vacuum expectation value of this expression in the normal ordering regularization scheme leads to the following expression of the background energy density of the field
\beq\label{eq:rhopsi1}
\rho_{\psi} \;=\; \frac{2}{a^4}\sum_{s=\pm}\int \frac{\diff^3\bp}{(2\pi)^3}\, \omega_p\,n_p^{(s)}\,,
\eeq
where the occupation numbers are given by
\beq\label{eq:np1}
n_p^{(s)} \;=\; \frac{1}{2}\left| \sqrt{1-\frac{a(m_{\psi}+y\phi)}{\omega_p}} e^{-is\theta}\, U_1^{(s)} - \sqrt{1+\frac{a(m_{\psi}+y\phi)}{\omega_p}} U_2^{(s)} \right|^2\,,
\eeq
(for details see e.g.~\cite{Greene:2000ew,Nilles:2001fg,Adshead:2018oaa}). With respect to the energy density of a scalar field, we notice two differences. First, we get a sum over the two polarizations. Furthermore, the factor two in front appears because a massive Dirac vector has the particle and antiparticle component.

\subsection{Inflaton fragmentation}\label{sec:frag}

\begin{figure*}[!t]
\centering
    \includegraphics[width=\textwidth]{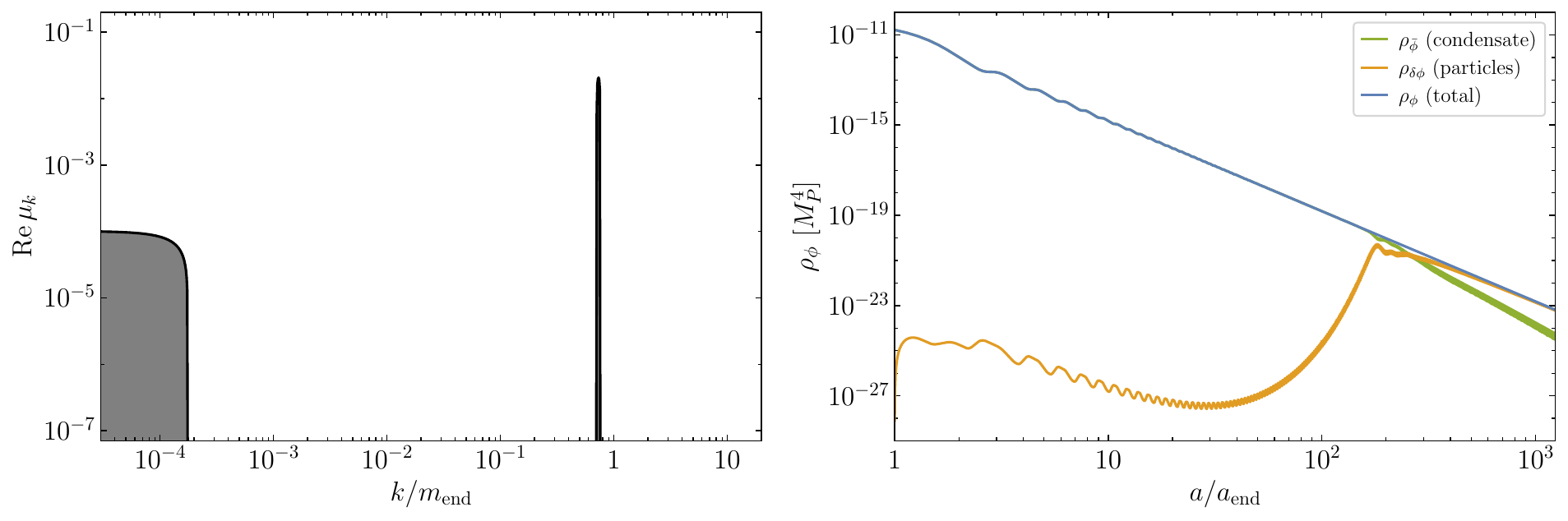}
    \caption{Left: Floquet chart for Hill's equation (\ref{eq:hills}). The gray bands show the regions of exponential growth of inflaton fluctuations. Right: Time dependence of the total inflaton energy density (blue), the energy density of the inflaton inhomogeneities (orange), and the energy density of the homogeneous component (green) for the quartic T-model (\ref{eq:Tquart}). For $a/a_{\rm end}<10^2$ these energies are computed using the linear approximation, while at later times lattice methods are utilized.}
    \label{fig:frag}
\end{figure*}

In a quartic potential, the self interaction of the inflaton field, although small in coupling value (c.f.~Eq.~(\ref{eq:lambda})), is sufficient to drive the resonant growth of its homogeneities $\delta\phi(t,\bx)$. Therefore, unless reheating is rapidly completed, the coherent oscillation approximation eventually ceases to be valid. At linear order the inflaton fluctuations satisfy the equation of motion
\beq\label{eq:eqdeltaphi}
\ddot{\delta\phi} + 3H\dot{\delta\phi} - \frac{\nabla^2\delta\phi}{a^2} + 12\lambda\phi(t)^2\delta\phi\;=\;0\,.
\eeq
Introducing the canonically normalized field\footnote{Here the ladder operators satisfy commutation relations, $[\boldsymbol{A}_{\bk},\boldsymbol{A}_{\bk'}^{\dagger}]=\delta^{(3)}(\bk-\bk')$, $[\boldsymbol{A}_{\bk},\boldsymbol{A}_{\bk'}]=[\boldsymbol{A}^{\dagger}_{\bk},\boldsymbol{A}^{\dagger}_{\bk'}]=0$, and the mode functions fulfil the Wronskian condition $X_kX_k^{*\,\prime}-X_k^*X_k'=i$.}
\begin{align}
X(\tau,\bx)\;&\equiv\; a(\tau)\,\delta\phi(\tau,\bx)\\
&=\; \int \frac{\diff^3\bk}{(2\pi)^3}\,e^{-i\bk\cdot\bx}\left(X_k(\tau)\boldsymbol{A}_{\bk} + X_k^*(\tau)\boldsymbol{A}_{-\bk}^{\dagger}\right)\,,
\end{align}
and the dimensionless time variable $z\equiv m_{\rm end}(\tau-\tau_{\rm end})$, Eq.~(\ref{eq:eqdeltaphi}) can be equivalently written as the (infinite) set of equations
\beq\label{eq:hills}
\frac{\diff^2 X_k}{\diff z^2} + \left[ \left(\frac{k}{m_{\rm end}}\right) + \mathrm{sn}^2\left(\frac{z}{\sqrt{6}},-1\right)\right] \;\simeq\;0\,,
\eeq
where the approximate equality indicates that we have disregarded terms that decrease rapidly after the onset of reheating (for details see e.g.~\cite{Garcia:2023eol}). The resulting equations are a particular case of Hill's equation, and for certain values of $k$ will present parametric resonance. The solutions can be written in general as~\cite{Magnus2004-br}
\beq
X_k(z) \;=\; e^{\mu_k z}g_1(z) + e^{-\mu_k z}g_2(z)\,,
\eeq
where $g_{1,2}$ are periodic functions on $z$, and the Floquet exponent $\mu_k$ is in general a complex quantity. The resonance is manifested as the exponential growth of those mode functions for which ${\rm Re}\,\mu_k\neq 0$. The left panel of Fig.~\ref{fig:frag} shows the Floquet chart for the quartic inflaton potential, where two main resonance bands can be identified, one at small momentum $k/m_{\rm end}\lesssim 2\times 10^{-4}$, and another one at $0.71\lesssim k/m_{\rm end}\lesssim 0.76$. The later, although narrow, has a larger Floquet exponent, and is the main driver for the exponential growth of the inflaton inhomogeneities.

In the linear approximation, the energy density of $\delta\phi$ can be computed as~\cite{Kofman:1997yn,Garcia:2021iag}
\beq
\rho_{\delta\phi} \;=\; \frac{1}{2(2\pi)^3a^4}\int \diff^3\bk\,\left|\omega_{k}X_k-iX_k'\right|^2\,,
\eeq
where the effective frequency is given by 
\beq
\omega_k^2 \;=\; k^2 - \frac{a''}{a} + 12\lambda\phi^2a^2\,.
\eeq
This density can be appreciated in the right panel of Fig.~\ref{fig:frag}, as the orange curve. At early times its magnitude is subdominant compared to that of the condensate (green curve). Nevertheless, the eventual exponential growth of the inhomogeneities is manifest, and it can be tracked using the linear approximation until it becomes comparable in magnitude to the condensate energy density. At this point, the linear approximation fails, as re-scatterings, manifested as couplings between mode functions of different $k$, become important. Large spatial gradients are generated, and the inflaton becomes {\em fragmented}. In this highly non-linear regime lattice methods to solve the PDE
\beq
\ddot{\phi}+3H\dot{\phi} - \frac{\nabla^2\phi}{a^2} + \partial_\phi V \;=\; 0\,,
\eeq
are favoured, such as \texttt{$\mathcal{C}\text{osmo}\mathcal{L}\text{attice}$}~\cite{Figueroa:2020rrl,Figueroa:2021yhd}. As Fig.~\ref{fig:frag} shows, eventually the condensate component, computed in this regime as the energy density of the spatially averaged field $\overline{\phi(t,\bx)}$
\beq
\rho_{\bar{\phi}}\;=\; \frac{1}{2}\bar{\dot{\phi}}^2 + V(\bar{\phi})\,,
\eeq
is replaced by a collection of free inflaton quanta, with energy density
\beq
\rho_{\delta\phi}=\rho_{\phi}-\rho_{\bar{\phi}}\,.
\eeq
Here the total energy density corresponds to
\beq
\rho_{\phi} \;=\; \overline{\frac{1}{2}\dot{\phi}^2 + \frac{1}{2a^2}(\nabla\phi)^2+V(\phi)}\,.
\eeq

\begin{figure*}[!t]
\centering
    \includegraphics[width=0.90\textwidth]{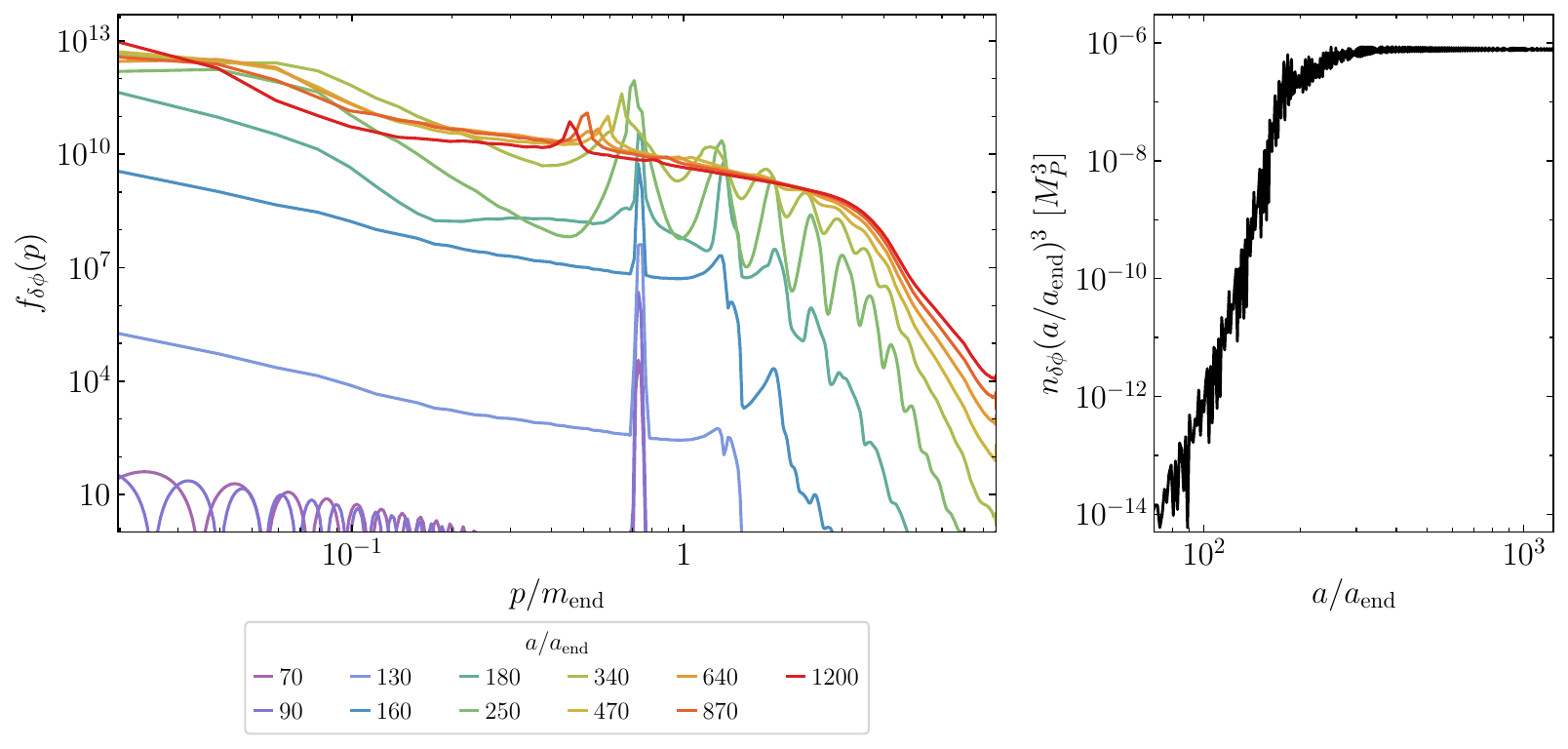}
    \caption{Left: The phase space distribution (PSD) for the inflaton inhomogeneities $\delta\phi$ evaluated at a selection of scale factors, during and after the fragmentation process. Right: The comoving number density of the inflaton fluctuations as a function of the scale factor. We assume here that reheating ends at $a_{\rm reh}\gg 10^{3}a_{\rm end}$.
    }
    \label{fig:phiPSD}
\end{figure*}

Of equal importance to the energy density is the PSD (or occupation number) of the inflaton field during and after fragmentation. If reheating is not completed prior to fragmentation, it will be the decay of the resonantly produced inflaton quanta what will be responsible for reheating the universe. In the Boltzmann approximation, their PSD and the decay amplitude are the inputs necessary to determine the production rate of daughter particles, and therefore the efficiency of reheating. Under the assumption that quantum statistical and in-medium effects (other than the induced inflaton mass) are negligible, the continuity equations for the decay of the fragmented inflaton take the form~\cite{Garcia:2023eol,Garcia:2023dyf}
\begin{align}\label{eq:contfrag1}
\dot{\rho}_{\delta\phi} + 3H(1+w_{\phi})\rho_{\delta\phi} \;&=\; -\Gamma_{\delta\phi}m_{\phi}n_{\delta\phi}\,,\\ \label{eq:contfrag2}
\dot{\rho}_{R} + 4H \rho_{R} \;&=\; \Gamma_{\delta\phi}m_{\phi}n_{\delta\phi}\,,
\end{align}
where $\rho_R$ denotes the energy density of the relativistic decay products, and 
\beq\label{eq:ndeltaphi}
n_{\delta\phi} \;=\; \int \frac{\diff^3\bp}{(2\pi)^3}\,f_{\delta\phi}(p)\,,
\eeq
denotes the inflaton quanta number density, obtained from integration of its PSD $f_{\delta\phi}(p)$, which must be determined numerically, as shown in Fig.~\ref{fig:phiPSD}.

\section{Fermion production before fragmentation}\label{eq:beforefrag}

In Section~\ref{sec:infosc}, we detailed the field content of our model and the corresponding dynamical equations to solve in order to determine the efficiency of reheating, assuming that the self-fragmentation of the inflaton has not occurred. In the present Section, we will focus on the computation of the fermion energy density in this regime, in order to determine under what conditions is reheating possible before inflaton fragmentation, if at all. To fully illustrate the role of short time-scale dynamics and quantum statistics, we will approach this analysis first from the Boltzmann perspective. Subsequently, we compare it to the full solution of the field equations in the Heisenberg picture. With this, we aim to identify if and when the perturbative picture breaks down.

\subsection{The Boltzmann approximation}\label{sec:boltzmann}

We now tackle the production of fermions from inflaton decay in the quartic potential (\ref{eq:Tquart2}) by means of the Boltzmann equation. As stated above, we consider here the direct dissipation of the inflaton condensate, without considering the presence of inflaton fluctuations $\delta\phi$. We will account for their effect on $\psi$-production in Section~\ref{sec:afterfrag}. 

Following~\cite{Ichikawa:2008ne,Nurmi:2015ema,Garcia:2020wiy,Garcia:2022vwm}, the Boltzmann equation for the fermion PSD in the presence of a coherently oscillating inflaton background reads
\begin{equation}
\begin{split}
   \frac{\partial f_{\psi}}{\partial t}-H|\boldsymbol{P}| \frac{\partial f_{\psi}}{\partial|\boldsymbol{P}|}\,=\,\frac{1}{P^{0}} &\sum_{n=1}^{\infty} \int  \frac{\diff^{3} \boldsymbol{K}_{n}}{(2 \pi)^{3} n_{\phi}} \frac{\diff^{3} \boldsymbol{P}^{\prime}}{(2 \pi)^{3} 2 P^{\prime 0}}(2 \pi)^{4} \delta^{(4)}\left(K_{n}-P-P^{\prime}\right)\left|\overbar{\mathcal{M}_{n}}\right|^{2} \\ & \times\left[f_{\phi}\left(K_{n}\right)\left(1-f_{\psi}(P)\right)\left(1-f_{\psi}\left(P^{\prime}\right)\right)-f_{\psi}(P) f_{\psi}\left(P^{\prime}\right)\left(1+f_{\phi}\left(K_{n}\right)\right)\right] . 
\label{eq:Boltzmannfordeltaphi}
\end{split}
\end{equation}
Here the physical four-momenta is denoted by uppercase letters, with $P$ and $P'$ corresponding to the momenta of the produced fermions. The condensate contributes to the collision term with one decay amplitude per harmonic oscillation mode, as per the Fourier decomposition of the quasi-periodic function $\mathcal{P}(t)$ in (\ref{eq:oscillating}),
\begin{equation}
\mathcal{P}(t) \;=\; \sum_{n=-\infty}^{\infty} \mathcal{P}_n e^{-in\omega_{\phi} t} \,.
\end{equation}
These modes behave as quasi-particles with momenta $K_n = \left(E_n, \bf{0} \right)$, where $E_n = n \, \omega_{\phi}$ is the energy of the $n^\text{th}$ mode. The frequency of oscillation for the quartic potential is related to the inflaton mass by
\beq \label{eq:omegaphi}
\omega_{\phi} \;=\; m_{\phi}\sqrt{\frac{2\pi}{3}}\frac{\Gamma(3/4)}{\Gamma(1/4)} \;\equiv\; c\,m_{\phi}\,.
\eeq
Here, $\mathcal{M}_{n}$ represents the transition amplitude corresponding to the production of a pair of fermion/anti-fermion $| f\rangle=| \bar \psi \psi\rangle$ from the vacuum by the oscillating inflaton background field, determined from
\begin{equation}
\big|\big\langle f \big| i \int \diff^{4} x \ \mathcal{L}_{I} \big| 0 \big\rangle \big|^{2}=\operatorname{Vol}_{4} \sum_{n=-\infty}^{\infty}\left|\overline{\mathcal{M}_{n}}\right|^{2}(2 \pi)^{4} \delta^{(4)}\left(K_{n}-P-P^{\prime}\right),
\label{eq:matrixelement}
\end{equation}
where Vol$_4$ denotes the space-time volume. The matrix elements $\mathcal{M}_{n}$ are found to be
\begin{equation}
\left|\overline{\mathcal{M}_{n}}\right|^{2}= \dfrac{2 n^2 \omega_\phi^2}{g_\psi} \bar{y}_n^2 \beta_n^2 \phi_{0}^{2}\,|\mathcal{P}_{n}|^{2} ,
\end{equation}
where we averaged over possible spin states. To account for the scalar and pseudoscalar couplings we have introduced the effective coupling
\begin{equation}
    \bar y_n(t) \, \equiv \, \sqrt{y^2+y_5^2/\beta_n^2(t)} \,, 
\end{equation}
where 
\beq
\beta_n(t) \, \equiv \,   \sqrt{1-\dfrac{4 m_{\psi,{\rm eff}}^2(t)}{n^2 \omega_\phi^2(t)} }  \,,
\eeq
is the kinematic phase space factor. The effective mass of the fermion, $m_{\psi,{\rm eff}}(t)=m_{\psi} + y\phi(t)+ i y_5 \phi(t)  \gamma_5 $, includes the contribution from the bare mass and the instantaneous expectation value of the inflaton. In what follows, unless explicitly stated, we will assume that the bare mass of the fermion can be neglected, $m_{\psi}\ll m_{\phi}$. This allows us to introduce the mass ratio 
\begin{equation}\label{eq:Rdef}
\mathcal{R} \equiv \, \left. \dfrac{4 m_{\psi,{\rm eff}}^2(t)}{\omega_\phi^2(t)} \right|_{\phi \rightarrow \phi_0} \; = \frac{1}{2\pi \lambda}\left(\frac{\Gamma(1/4)}{\Gamma(3/4)}\right)^2\left(\frac{m_{\psi,\mathrm{eff}}}{\phi_0}\right)^2 = \frac{12}{2 \pi}\left(\frac{\Gamma(1/4)}{\Gamma(3/4)}\right)^2 \frac{m^2_{\psi,\mathrm{eff}}}{m^2_{\mathrm{end}}}\left(\frac{a}{a_{\mathrm{end}}}\right)^2\,.
\end{equation}
In terms of the couplings we can rewrite the previous expression as
\begin{equation}\label{eq:Rdefconst}
    \mathcal{R}  = \; \dfrac{\bar{y}^2 \Gamma^2(1/4)}{2\pi \lambda \Gamma^2(3/4)} \;\simeq\; 0.4 \, \left( \dfrac{\bar{y}}{10^{-6}}\right)^2 \left( \dfrac{3.3 \times10^{-12}}{\lambda}\right)\,,
\end{equation}
which reduces to a constant (only for a quartic potential). The kinematic factor $\beta_n$ can in turn be expressed as
\begin{equation}\label{eq:betansimp}
  \beta_n  = \sqrt{1-\dfrac{\mathcal{R} \mathcal{P}^2 }{n^2 } } \,.
\end{equation}
If the decay of an oscillating mode into a pair of fermions is kinematically blocked at the beginning of reheating, it will remain so, setting a threshold for particle production. The presence of $\mathcal{P}(t)$ in (\ref{eq:betansimp}) shows that there is not a full kinematic suppression, as the phase space can open during a fraction of the oscillation of the inflaton. Roughly speaking, $\mathcal{R}\gtrsim 1$ signals the dependence of the particle production process on short time scales, and therefore the relevance of parametric resonance effects~\cite{Garcia:2021iag}. 

The last necessary ingredient in order to integrate the Boltzmann equation (\ref{eq:Boltzmannfordeltaphi}) corresponds to the inflaton PSD, which for the spatially homogeneous condensate corresponds to the zero mode,
\begin{equation}
f_{\phi}(\boldsymbol{K}, t)=(2 \pi)^{3} n_{\phi}(t) \delta^{(3)}(\boldsymbol{K})\,.
\end{equation}
For our perturbative analysis we will disregard the relevance of inverse decays and Pauli blocking. Although these effects can be taken into account numerically in the Boltzmann approximation, as we will show in Section~\ref{sec:afterfrag}, we will obtain the full solution of the dynamics in the Heisenberg picture in the next sub-Section. Our goal here is to emphasize the shortcomings of the aforementioned approximations. With this in mind, we can rewrite the Boltzmann equation (\ref{eq:Boltzmannfordeltaphi}) as 
\begin{align} \notag
\frac{\partial f_{\psi}}{\partial t}\;-\;H|\boldsymbol{P}| \frac{\partial f_{\psi}}{\partial|\boldsymbol{P}|}  \;&\simeq \; \sum_{n=1}^{\infty} \frac{1}{P^{0}} \int \frac{\diff^{3} \boldsymbol{K}_{n}}{(2 \pi)^{3} n_{\phi}} \frac{\diff^{3} \boldsymbol{P}^{\prime}}{(2 \pi)^{3} 2 P^{\prime 0}}(2 \pi)^{4} \delta^{(4)}\left(K_{n}-P-P^{\prime}\right)\left|\overline{\mathcal{M}_{n}}\right|^{2} f_{\phi}\left(K_{n}\right) \nonumber\\  \notag
& = \; 2 \pi  \sum_{n=1}^{\infty} \frac{\left|\overline{\mathcal{M}_{n}}\right|^{2}}{n^{2} \omega_\phi^{2} \beta_{n}} \delta\left(|\boldsymbol{P}|-\frac{1}{2} n \omega_\phi \beta_{n}\right)\\
& = \dfrac{4 \pi }{g_\psi} \phi_0^2 \sum_{n=1}^\infty \bar y_n^2 \beta_n |\mathcal{P}_n|^2 \delta\left(|\boldsymbol{P}|-\frac{1}{2} n \omega_\phi \beta_{n}\right)\,.
\label{eq:Boltzmannsimp}
\end{align}
The independence of the right-hand side of the equation above with respect to $f_{\psi}$ allows for a straightforward integration,
\begin{equation}\label{eq:solboltzt}
f_{\psi}(|\boldsymbol{P}|, t) \; \simeq \; \dfrac{4 \pi }{g_\psi}  \sum_{n=1}^{\infty} |\mathcal{P}_n|^{2}\int_{t_{\mathrm{end}}}^{t} \diff t^{\prime} \bar y_n^2\left(t^{\prime}\right) \beta_n \left(t^{\prime}\right) \phi_{0}^{2}\left(t^{\prime}\right) \delta\left(\frac{a(t)}{a\left(t^{\prime}\right)}|\boldsymbol{P}|-\frac{1}{2} n  \beta_{n}(t^{\prime}) \omega_{\phi}(t^{\prime})\right) \,,
\end{equation}
assuming a vanishing population of fermions at the end of inflation (see e.g.~\cite{Ballesteros:2020adh}). Integration is simplified by switching the time variable from cosmic time to the scale factor. Defining the dimensionless comoving momentum
\beq
q \;\equiv\; \frac{a(t)|\boldsymbol{P}|}{a_\text{end} m_\text{end}}\,,
\eeq
the solution (\ref{eq:solboltzt}) may be rewritten as\footnote{A more accurate approximation, which accounts for the difference between the full T-model potential (\ref{eq:Tquart}) and the quartic approximation (\ref{eq:Tquart2}) at the end of inflation, consists in replacing $c\rightarrow \frac{a' \omega_\phi(a')}{a_{\mathrm{end}} m_{\mathrm{end}}}$ in (\ref{eq:fpsiint}).}
\begin{align}\notag \displaybreak[0]
f_{\psi}(|\boldsymbol{P}|, t) \;& \simeq \; \dfrac{4 \pi }{g_\psi}  \left( \dfrac{1}{a_\text{end} m_\text{end}}\right)\sum_{n=1}^{\infty} |\mathcal{P}_n|^{2}\int_{a_{\mathrm{end}}}^{a(t)} \diff a^{\prime} \dfrac{\phi_{0}^{2}\left(a^{\prime}\right) }{ H(a')} \bar y_n^2\left(a^{\prime}\right) \beta_n \left(a^{\prime}\right) \delta\left(q- \frac{1}{2}nc\beta_{n}(a') \right)\\ \label{eq:fpsiint}
& \simeq \; \dfrac{4 \pi }{g_\psi}  \left( \dfrac{\phi_{\rm end}^2}{a_\text{end} m_\text{end}H_{\rm end}}\right)\sum_{n=1}^{\infty} |\mathcal{P}_n|^{2}\int_{a_{\mathrm{end}}}^{a(t)} \diff a^{\prime}  \bar y_n^2\left(a^{\prime}\right) \beta_n \left(a^{\prime}\right) \delta\left(q- \frac{1}{2}nc\beta_{n}(a') \right) \,.
\end{align}
In the Boltzmann approximation short time-scale dependent quantities are interpreted in the oscillation-average sense, allowing for an adiabatic computation in perturbation theory~\cite{Garcia:2020wiy}. Applying this reasoning for the $\mathcal{P}$-dependent quantities inside the integral sign in (\ref{eq:fpsiint}), we finally obtain
\begin{equation}
\begin{aligned}
f_{\psi}(|\boldsymbol{P}|, t) \; &  \simeq \; \dfrac{4 \pi }{g_\psi}  \left( \dfrac{\phi_\text{end}^2 }{ m_\text{end} H_\text{end}}\right) \left( \dfrac{a(t)}{a_\text{end}}-1 \right) \sum_{n=1}^{\infty}  \bar y_n^2 \beta_n |\mathcal{P}_n|^{2} 
 \delta\left( q- \frac{1}{2}nc\beta_{n} \right)  \,.
 \label{eq:PSDnegbetaprime}
\end{aligned}
\end{equation}
As a result, the Boltzmann approach predicts a series of discrete peaks located at
\beq\label{eq:Boltzpeaks}
q_{\rm peak}^{(n)} \;=\; \frac{1}{2}nc\beta_n\,,\qquad (n \text{ odd})\,.
\eeq
The amplitude of each peak increases linearly with the scale factor and scales as $ \bar y_n^2\beta_n  |\mathcal{P}_n|^{2}$, overwhelmingly dominated by the contribution from the first oscillation mode. The position of the first five Boltzmann peaks can be appreciated in Figs.~\ref{fig:1e-8} and \ref{fig:1e-6}, compared to the exact distribution obtained by direct integration of the $\psi$ equations of motion. The previous result also shows that the efficiency of particle production is controlled by the effective coupling $\bar{y}^2/\lambda$~\cite{Greene:1998nh}. Therefore, by adequately re-scaling the inflaton-fermion couplings the results of our analysis can be generalized to inflaton potentials with a normalization different than (\ref{eq:lambda}).

\begin{figure*}[!t]
\centering
    \includegraphics[width=0.6\textwidth]{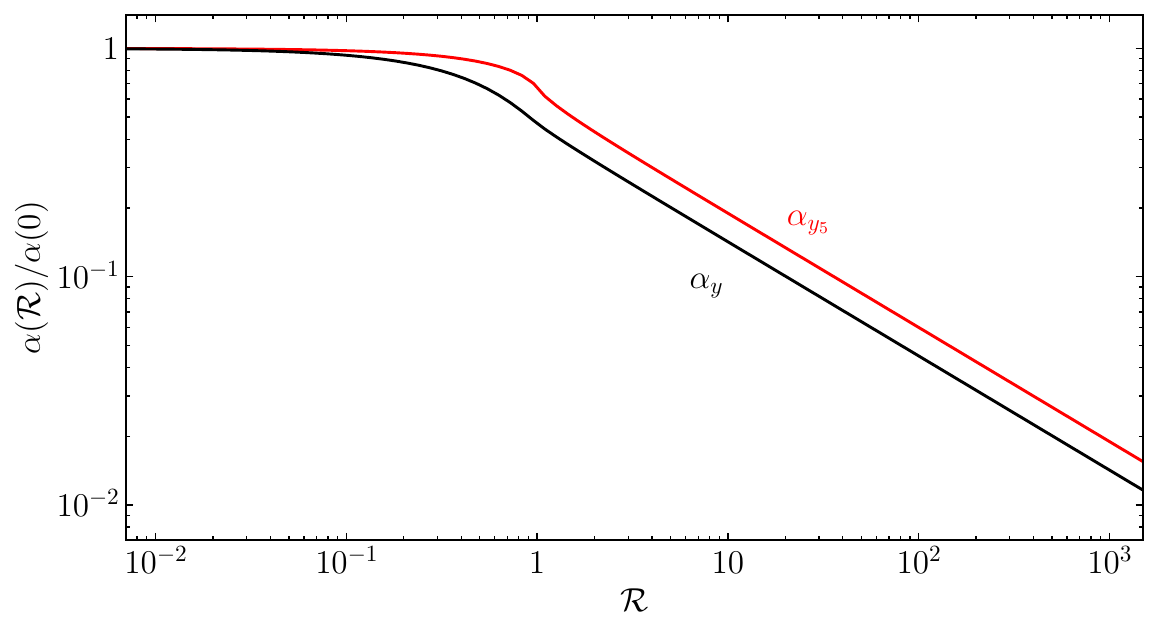}
    \caption{Kinematic suppression factor for the effective inflaton-fermion coupling (\ref{eq:yeff}), as a function of the mass ratio $\mathcal{R}$.
    }
    \label{fig:kinF}
\end{figure*}

The energy density of the produced fermions is determined by integrating over their physical momentum,
\begin{equation}
    \rho_\psi (t) \, = \, \dfrac{g_\psi}{(2\pi)^3}  \int \diff^3 \boldsymbol{P} \sqrt{|\boldsymbol{P}|^2+ m_\psi^2} f_\psi(\boldsymbol{P},t) \, .
\end{equation}
A straightforward computation yields
\begin{align}
    \rho_\psi (t) \; &\simeq \; 
 \dfrac{1}{4\pi} \left( \dfrac{\phi_\text{end}^2 m_\text{end}^3 }{  H_\text{end}}\right)   \left(  \dfrac{a(t)}{a_\text{end}} \right)^{-4} \left( \dfrac{a(t)}{a_\text{end}}-1 \right) \sum_{n=1}^{\infty}  c^3n^3 \bar y_n^2\beta_n^3   |\mathcal{P}_n|^{2} \\ \label{eq:rhopsiBoltz}
 &\equiv\; \frac{y_{\rm eff}^2}{\pi}\lambda^{1/4}\rho_{\rm end}^{3/4}M_P \left(  \dfrac{a(t)}{a_\text{end}} \right)^{-4} \left( \dfrac{a(t)}{a_\text{end}}-1 \right)\,,
\end{align}
where we have introduced the effective coupling~\cite{Garcia:2020wiy}
\beq\label{eq:yeff}
y_{\rm eff}^2 \;=\; 18\sum_{n=1}^{\infty} c^3n^3 \bar y_n^2\beta_n^3   |\mathcal{P}_n|^{2} \;\equiv\; \begin{cases}
\alpha_y (\mathcal{R})y^2\,,\\
\alpha_{y_5}(\mathcal{R})y_5^2\,.
\end{cases}
\eeq
In the absence of kinematic blocking $\beta\simeq 1$, $\alpha_y=\alpha_{y_5}\simeq0.51$. Fig.~\ref{fig:kinF} shows the dependence of the effective coupling on the kinematic mass ratio $\mathcal{R}$ showing the magnitude of the suppression on the particle production rate. For $\mathcal{R}\gg 1$, $\alpha_{y,y_5}\propto \mathcal{R}^{1/2}$.

Under the assumption of instantaneous thermalization the temperature of the relativistic decay products would be given by
\beq
T \;=\; \left(\frac{30 \rho_{\psi}}{\pi^2g_{\rm reh}}\right)^{1/4}\,,
\eeq
where $g_{\rm reh}$ denote the number of relativistic degrees of freedom during reheating. For $a\gg a_{\rm end}$, $T\propto a^{-3/4}$. If reheating was completed perturbatively before the onset of fragmentation, that is $\rho_{\psi}=\rho_{\phi}$, the end of reheating would occur for a scale factor
\beq
\frac{a_{\rm reh}}{a_{\rm end}} \;\simeq\; \frac{\pi \rho_{\rm end}^{1/4}}{y_{\rm eff}^2 \lambda^{1/4} M_P}\,,
\eeq
and a reheating temperature
\beq\label{eq:Trehpert}
T_{\rm reh}\;\simeq\; \left( \frac{30\lambda}{\pi^6 g_{\rm reh}}\right)^{1/4} y_{\rm eff}^2 M_P\,.
\eeq
We note that reheating before fragmentation corresponds to $a_{\rm reh}\lesssim 180\, a_{\rm end}$ (see Fig.~\ref{fig:frag}). This would require that $y_{\rm eff}\gtrsim 0.3$, value for which the perturbative Boltzmann approximation is expected to break down. This leaves us with two possibilities: (1) Reheating can be completed before fragmentation, but a non-perturbative computation is necessary to determine the fermion production rate. Or (2) reheating is completed after fragmentation from the decay of the free inflaton quanta. We explore both scenarios in the following (sub)sections, finding that the estimate (\ref{eq:Trehpert}) is inadequate for both.

\subsection{Non-perturbative dynamics}\label{sec:nonpert}

We now proceed to quantify the efficiency of post-inflationary fermion production by directly solving the Dirac equation (\ref{eq:Diraceq}) prior to the onset of inflaton fragmentation. One of our goals is to quantify the range of values of the Yukawa couplings for which the perturbative approach accurately describes pre-fragmentation fermion production. More specifically, we determine when non-perturbative production mechanisms become crucial in determining the instantaneous energy density $\rho_{\psi}$. Our second goal is to quantify when backreaction effects are important prior to fragmentation, which will allow us to estimate the strength of the couplings necessary for prompt inflaton decay.

The Boltzmann approach discussed in the previous sub-section accounts for the decay of the inflaton condensate via the tree-level amplitude, but fails to capture production from the expanding background (gravitational production) and from the short time-scale, non-adiabatic oscillations of the inflaton condensate around the minimum of its potential. In the Boltzmann picture we have also disregarded the effect of quantum statistics (Pauli blocking). Moreover, we cannot describe the excitations of modes with comoving momentum $p<a_{\rm end}H_{\rm end}$, as at the beginning of reheating they would be outside the horizon. Nevertheless, for sufficiently small couplings the Boltzmann and Heisenberg approaches agree. As we will discuss below, the range of agreement corresponds to couplings so small that reheating cannot happen before (Big Bang) nucleosynthesis (BBN)~\cite{Garcia:2023eol}, and hence successful reheating always happens well into the non-perturbative regime. For the non-perturbative approach, we numerically evolve the Dirac equation in the form~(\ref{eq:diraceq1}), and evaluate the occupation number and energy density as given in Eqs.~(\ref{eq:rhopsi1}) and(\ref{eq:np1}). We account for modes with $p<a_{\rm end}H_{\rm end}$ by beginning the integration a few e-folds before the end of inflation, in order to properly initialize horizon-exiting modes in the Bunch-Davies vacuum. 

\begin{figure}[t]
\centering
     \includegraphics[width=\textwidth]{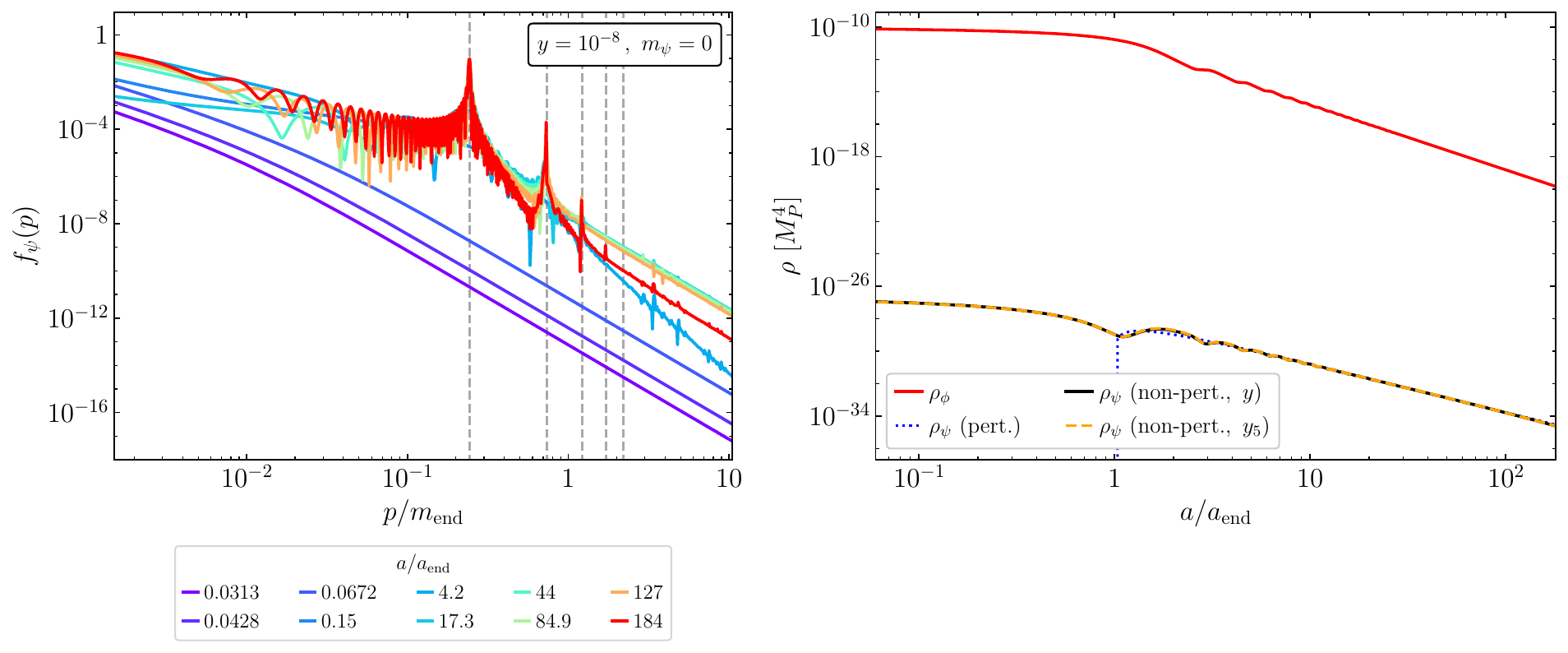}

           \caption{
            Left: Pre-fragmentation fermion PSD (\ref{eq:psdnp}) evaluated at different times (see legend) for the Yukawa coupling $y=10^{-8}$. The position of the first five peaks predicted by the Boltzmann approximation (\ref{eq:Boltzpeaks}) are shown as the vertical dashed gray lines. Right: Exact pre-fragmentation energy density of $\psi$ obtained from the integration of the PSD shown in the left panel (black, continuous), compared to the perturbative Boltzmann approximation (blue, dotted). For this case $\mathcal{R}\simeq0$ and kinematic suppression is negligible. For reference we also show the energy density of the inflaton condensate (red, continuous), and the energy density for the case of an axial coupling with strength $y_5=10^{-8}$ (dashed, orange).}
    \label{fig:1e-8}
\end{figure}

\begin{figure}[t]
\centering
     \includegraphics[width=\textwidth]{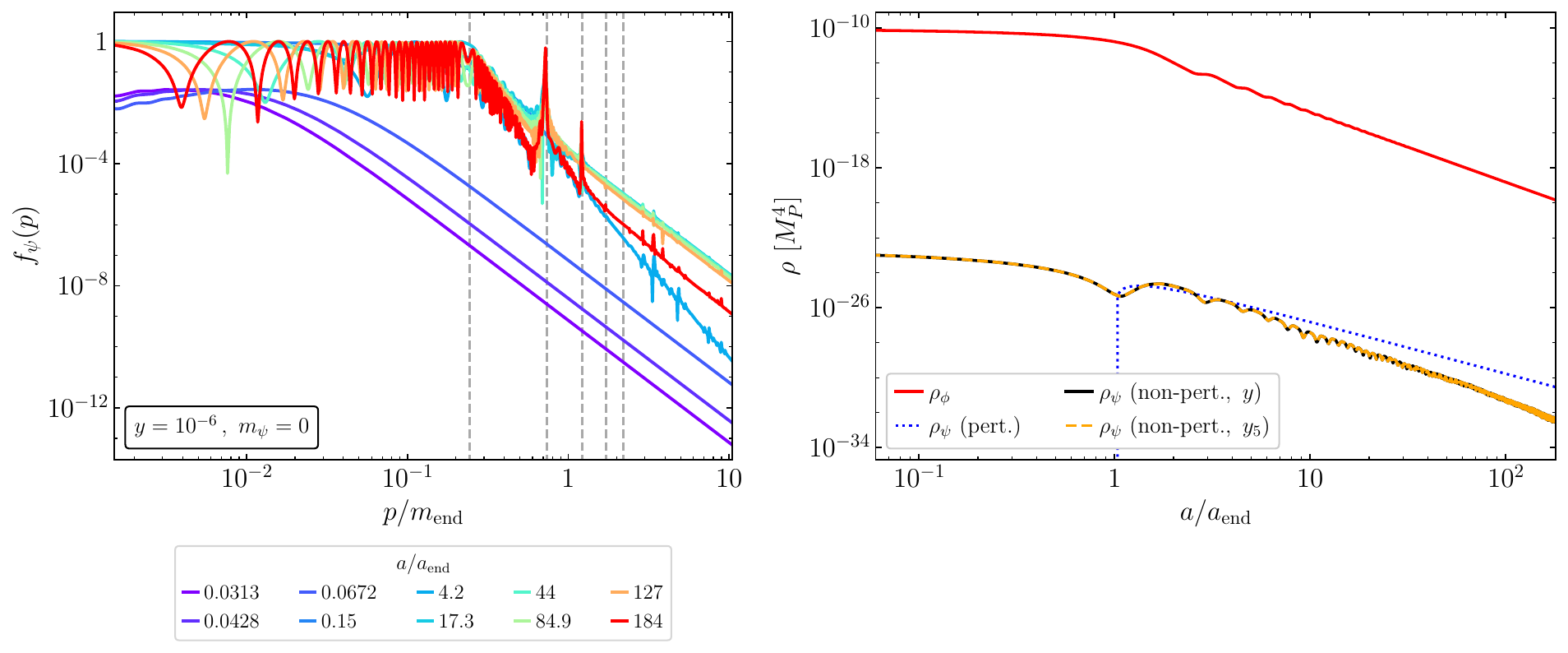}

            \caption{As in Fig.~\ref{fig:1e-8}, the fermion PSD before fragmentation, for a Yukawa coupling $y=10^{-6}$ (left), and a comparison of different approximations to $\rho_{\psi}$ (right).}
    \label{fig:1e-6}
\end{figure}

Our results in the small Yukawa coupling $y$ regime are summarized in Figs.~\ref{fig:1e-8}-\ref{fig:1e-6}, where for convenience we show in the left panels the PSD for the fermions $\psi$
\beq\label{eq:psdnp}
f_{\psi}(p) \;=\; \sum_{s=\pm}n_p^{(s)}\,,
\eeq
(c.f.~Eq.~(\ref{eq:np1})) for a selection of scale factors $a/a_{\rm end}\lesssim 184$, corresponding to pre-fragmentation dynamics. In these panels we also show as vertical dashed lines the position of the Dirac-delta peaks of the Boltzmann approximation (\ref{eq:Boltzpeaks}). The right panels of the aforementioned figures depict the evolution of the inflaton energy density (solid, red), the exact non-perturbative energy density $\rho_{\psi}$~(\ref{eq:rhopsi1}) (solid, black), and the perturbative Boltzmann approximation (\ref{eq:rhopsiBoltz}) (blue, dashed). We emphasize that for our non-perturbative (Heisenberg) computation we have not accounted for the backreaction of $\psi$-production into the inflaton dynamics. We will address this shortcoming in our upcoming discussion. For full comparison we also show the exact energy density in the case of a pseudoscalar coupling with strength $y_5=y$ (orange, dashed).

From our previous discussion of the kinematic mass ratio $\mathcal{R}$, we expect the perturbative Boltzmann approach to provide an accurate approximation to the reheating process for $y\lesssim 10^{-6}$ ($y^2/\lambda\lesssim 0.3$). Fig.~\ref{fig:1e-8} shows that, for $y=10^{-8}$, the perturbative and non-perturbative approaches provide equivalent results for the fermion energy density prior to fragmentation. We note that this is the case even when the non-perturbative computation also accounts for the evolution of the fermion modes during inflation. As predicted in the previous Section, the fermion energy density redshifts like matter, $\rho_{\psi}\propto a^{-3}$, after the first oscillation of the inflaton. In the absence of fragmentation this redshift would lead to a inflaton-fermion crossover at the reheating temperature $T_{\rm reh}\simeq 43$ MeV, just above the BBN lower bound. Nevertheless, in the presence of fragmentation, as we will discuss in the next section, the crossover temperature is significantly reduced, making this Yukawa coupling phenomenologically unsuitable. The PSD of $\psi$, presented in the left panel of Fig.~\ref{fig:1e-8}, is shown for a selection of scale factors starting from the end of inflation, coded by color. For the present coupling, we note that the Pauli limit $f_{\psi}=1$ is not saturated. The PSD clearly presents a series of peaks which coincide in position with the delta function peaks in the Boltzmann approach. 

\begin{figure}[t]
\centering
     \includegraphics[width=\textwidth]{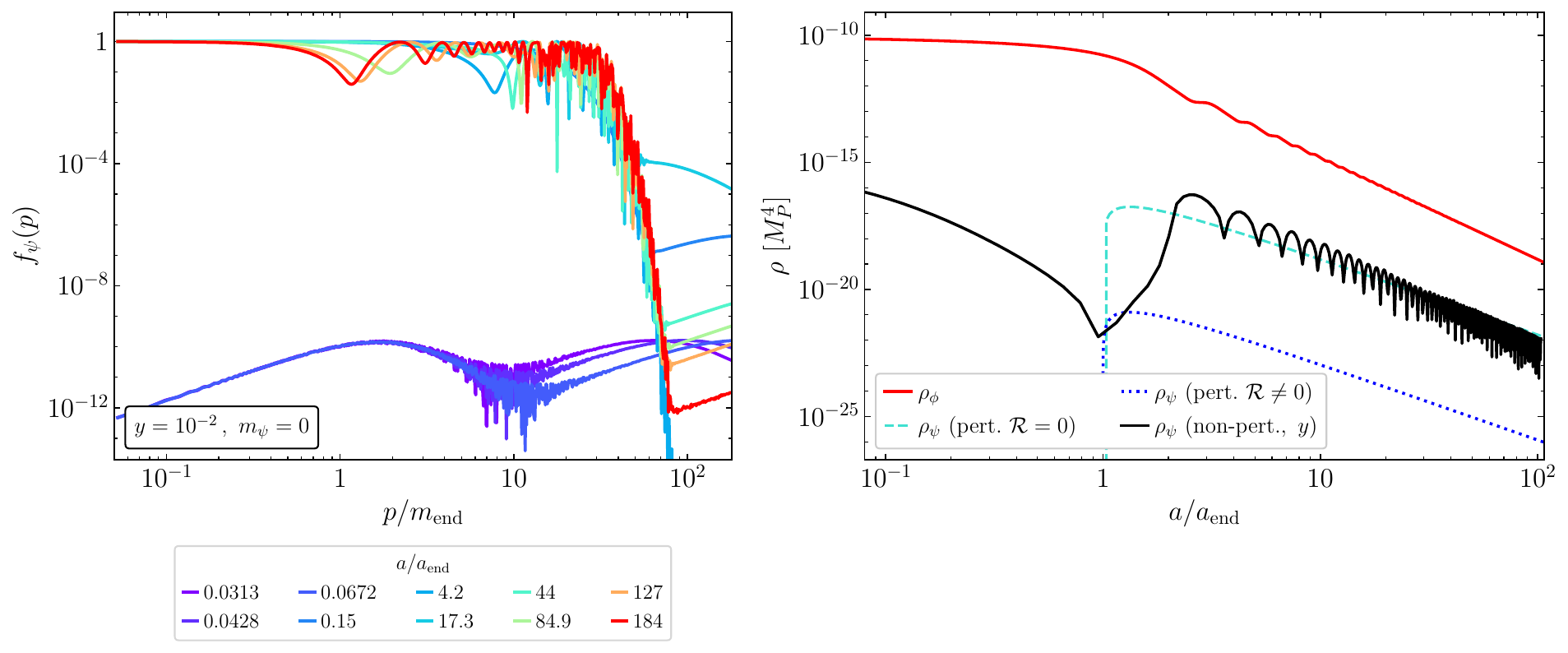}
                
            \caption{Left: As in Fig.~\ref{fig:1e-8}, the fermion PSD before fragmentation, for a Yukawa coupling $y=10^{-2}$. Right: Exact pre-fragmentation energy density of $\psi$ obtained from the integration of the PSD shown in the left panel (black, continuous), compared to the perturbative Boltzmann approximation not accounting for kinematic suppression, $\mathcal{R}=0$ (light blue, dashed), and accounting for kinematic suppression, $\mathcal{R}\neq0$ (dark blue, dotted). For reference we also show the energy density of the inflaton condensate (red, continuous). }
    \label{fig:1e-2}
\end{figure}

For both couplings discussed above the PSD shows a UV scaling law, $f_{\psi}\propto q^{-4}$ for $q\gtrsim H_{\mathrm{end}}/m_{\mathrm{end}}$. This behaviour can be understood from the parameter combination governing non-adiabaticity. The occupation number, $n_p$, is proportional to the small parameter $B \propto \frac{(a m)'}{p^2}$~\cite{Peloso:2000hy}. Consequently, $n_p^2= |B|^2 \propto p^{-4}$, as seen in the left panel of Figs \ref{fig:1e-8} and \ref{fig:1e-6}. For momenta $p/m_{\rm end}\gtrsim 40$ the PSD becomes exponentially suppressed, ensuring the convergence of the number and energy densities. The $f_{\psi}\propto q^{-4}$ scaling law is observed for sub-horizon modes for Yukawas up to $y\sim10^{-4}$, beyond which the power law changes to an exponentially decaying UV tail, as seen for example in the left panel of Fig.~\ref{fig:1e-2}. An intuitive explanation for this change from a scaling law to an exponential suppression is that while IR modes, which are initially super-horizon, can only be accounted for non-perturbatively, the UV modes are generically perturbative and seem to, at the level of the PSD, be in this regime up to larger couplings. However, the precise value of the coupling at which the perturbative nature of the PSD modes breaks down can not be determined analytically and must instead be inferred from a numerical computation.

By increasing the coupling we find that the effects of the short time-scale oscillations and Pauli blocking become significant for Yukawa couplings as small as $y=10^{-7}$ ($y^2/\lambda\simeq 3\times 10^{-3}$). Fig.~\ref{fig:1e-6} shows the results of the exact evaluation of PSD and energy densities for the coupling $y=10^{-6}$. For this coupling, kinematic effects in the Boltzmann approximation are still negligible, which is why we show only one perturbative curve in the right panel. Nevertheless, the exact energy density follows only the perturbative prediction only during the first couple of oscillations of the inflaton. Afterwards, the black (and dashed) curve is reduced in magnitude with respect to the dotted line, amounting to a relative suppression of approximately two orders of magnitude upon the onset of fragmentation. Switching to the left panel, we note that such a suppression is originated from the saturation of the PSD for the mode at the main Boltzmann peak, and smaller momenta. The spin 1/2 fermions cannot therefore continue being produced close to the main peak, and instead must rely on the population of higher momentum modes, which are less efficiently excited.

From the previous results we note that the perturbative Boltzmann approximation cannot accurately describe the reheating process for the range of couplings that would allow a decay of $\phi$ before its eventual fragmentation. Moreover, the Boltzmann approximation also fails to describe the evolution of the fermion PSD in the range of couplings that would allow the inflaton to decay prior to BBN after the onset of fragmentation (see (\ref{eq:fragTreh}) below). We therefore conclude that, in the absence of prompt thermalization that could redistribute the $\psi$ energy density into the Standard Model degrees of freedom and alleviate the Pauli suppression, the Boltzmann approximation fails to adequately describe the reheating process for our model in the phenomenologically viable range of couplings. It is therefore indispensable to study the dynamics of the system by non-perturbative means.\\

Having addressed small couplings, we proceed now to explore large couplings to analyze the possibility of pre-fragmentation reheating. Fig.~\ref{fig:1e-2} shows the result of increasing the Yukawa coupling up to $y=10^{-2}$. The left panel shows the non-perturbatively computed PSD. Comparison to Fig.~\ref{fig:1e-6} shows that the range of momenta has been shifted to larger values, as now the Pauli limit for the occupation numbers has been reached for $p/m_{\rm end}\lesssim 30$. Ignoring the kinematic factor in (\ref{eq:Boltzpeaks}), this would correspond to the saturation of the first 60 Boltzmann modes $q_{\rm peak}^{(n)}$. For larger momenta a sharp decrease in the PSD is observed. The right panel of Fig.~\ref{fig:1e-2} shows the exact energy density $\rho_{\psi}$ compared to two perturbative approximations. The first one (light blue, dashed) ignores the effect of kinematic blocking in the Boltzmann approximation. We note that, despite the fact that neither the maximum of $\rho_{\psi}$ nor its slope coincide with the exact result, the magnitude at the onset of fragmentation is coincidentally similar. The second perturbative curve (dark blue, dotted) accounts for kinematic blocking. Notably, what should be a more accurate description of the particle production process relative to the dashed curve, actually underestimates the exact result by about four orders of magnitude. This behavior is in contrast with fermionic preheating in a quadratic potential~\cite{Garcia:2021iag}, and stems from the availability of a continuous range of momenta to be saturated, as opposed to the discrete set predicted by the Boltzmann result. It is left for future endeavours to determine if, driven by interactions that we neglect here, a rapid thermalization of $\psi$ with Standard Model degrees of freedom would alleviate the Pauli suppression and allow for the pre-fragmentation decay of the inflaton at Yukawa couplings as large as these.
\begin{figure}[t]
    \centering
    \includegraphics[width=0.56\linewidth]{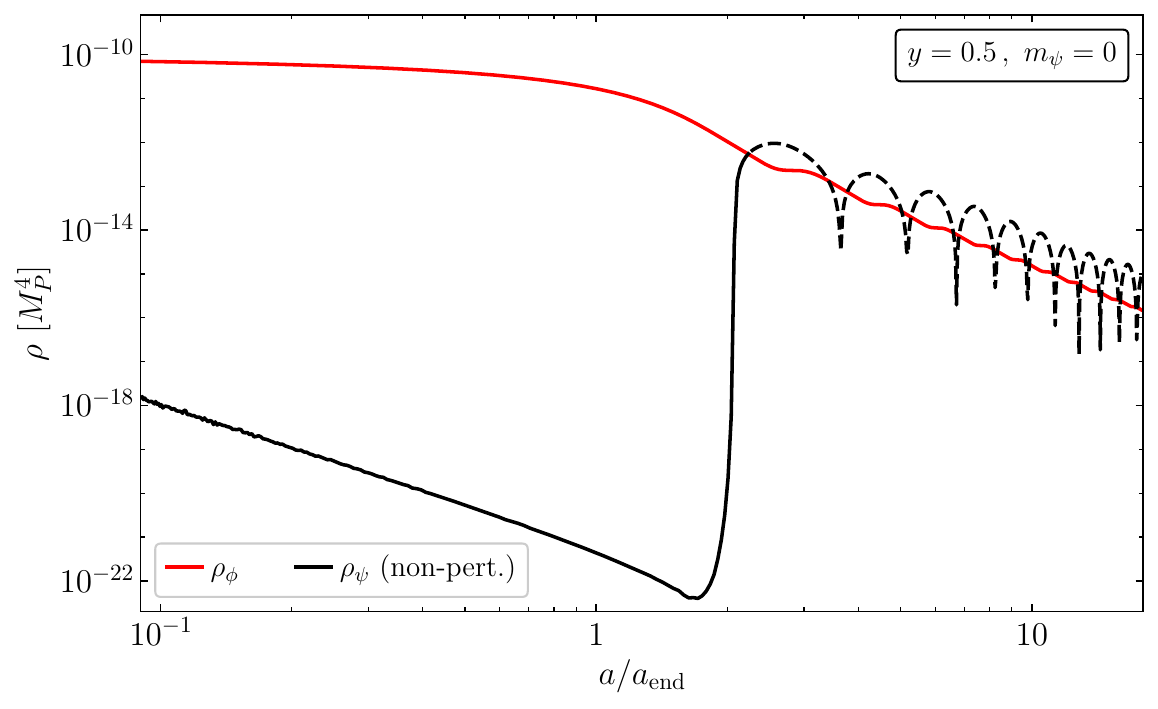}
        \caption{Inflaton (red) and non-perturbative fermion (black) energy densities as a function of the scale factor for a Yukawa coupling, $y=0.5$. Ignoring backreaction effects on the dynamics of $\phi$, reheating would be achieved in the first oscillation of the inflaton condensate.}
    \label{fig:y=0.5_energy_density}
\end{figure}

Purely within the context of the model defined by the action (\ref{eq:action}), we have to go to couplings as large as $y\gtrsim 0.2$ to obtain an energy density for the fermions comparable to $\rho_{\phi}$ prior to fragmentation. Fig.~\ref{fig:y=0.5_energy_density} shows one of such examples, for a coupling $y=0.5$. There, equality in the energy densities is achieved within the first oscillation of the inflaton, suggesting a nearly instantaneous completion of reheating. Nevertheless, in obtaining this result we have ignored the backreaction of the sudden growth of $\psi$ into the inflaton field. This neglect applies both to the dissipative dynamics, as Fig.~\ref{fig:y=0.5_energy_density} does not show the depletion of $\rho_{\phi}$ required by energy-momentum conservation, but also at the fluctuation level. For such large Yukawa couplings, re-scattering effects cannot be ignored, and a large amount of inflaton inhomogeneities may be generated. Backreaction effects in an inflaton-fermion system have been discussed in the context of axion inflation~\cite{Adshead:2018oaa,Mirzagholi:2019jeb,Adshead:2022ecl}, but not extensively in the context of fermion preheating (see however~\cite{Baacke:1998di,Berges:2010zv}). Such an exploration is beyond the scope of this work. 
    
    We finish this section by noting that our analysis has been limited to tree-level interactions between the inflaton and fermion sectors. In a quantum theory, these interactions will also be manifested as corrections to the scalar potential, therefore modifying the background dynamics and consequently the efficiency of reheating. As is well known, inflationary T-models are naturally motivated within supersymmetric theories, in particular in no-scale supergravity~\cite{Kallosh:2013hoa}. In such a set-up, the tadpole and quadratic radiative divergences of the effective potential would naturally be cancelled. In our present scenario with only an inflaton field and a fermion $\psi$, we can assume that these contributions are tuned to zero. Nevertheless, a correction of the Coleman-Weinberg form must be considered, which at the one-loop order is given by~ \cite{Ellis:2025bzi, Han:2025cwk, Kazakov:2023tii}
    \begin{equation}
    \Delta V(\phi) \;=\; -\frac{y^4\phi^4}{32\pi^2}\left[ \ln\left(\frac{y^2\phi^2}{\mu^2}\right) - \frac{3}{2} \right]\,,
    \end{equation}
    where $\mu$ denotes the renormalization scale. The dependence of the correction on this parameter is weak, and can be removed at the one-loop order by considering a renormalization group-improved effective potential allowing also for the running of the coupling constant $y$~\cite{Ellis:2025bzi}. Focusing on the prefactor, we observe that $\Delta V(\phi)$ will be equally important for the background evolution as well as for the tree-level potential for $y\gtrsim (32\pi^2\lambda)^{1/4}\simeq 10^{-2}$. We conclude that, in the strong backreaction regime, which we found to be necessary in order to successfully reheat the universe, radiative corrections cannot be ignored, further complicating the determination of the reheating energy scale. Not quantifying these effects represents the major shortcoming of our analysis, which we plan to revisit in future work (see however, e.g.~\cite{Gross:2015bea,Ellis:2025bzi,Alexandre:2025ixz}). 
    
In summary, we find that, at tree-level the transfer of an $\mathcal{O}(1)$ fraction of the inflaton energy density into fermions before the onset of fragmentation is only possible for $y\gtrsim 0.2$ ($y^2/\lambda\gtrsim 10^{10}$). For smaller values, the self-interaction of $\phi$ will resonantly source inflaton homogeneities faster than the inflaton can decay. The post-fragmentation production of fermions will be discussed in the following section.

\section{Fermion production after fragmentation}\label{sec:afterfrag}

For Yukawa couplings $y\lesssim 0.2$, the fragmentation of the inflaton occurs earlier than the completion of reheating for T-model potential (\ref{eq:Tquart}). Once this process occurs, the inflaton energy density is dominated by its inhomogeneities, which can be interpreted as free quanta with the PSD shown in Fig.~\ref{fig:phiPSD}. These quanta depend on the rapidly redfshifting condensate component to acquire a non-vanishing mass (at tree-level) and be kinematically allowed to decay into $\psi$. As we discussed in Section~\ref{sec:frag}, assuming quantum statistics, in medium effects, and short-time scale dynamics do not play a role in the decay process, the post-fragmentation reheating process can be described upon the solution of the continuity equations (\ref{eq:contfrag1})-(\ref{eq:contfrag2}). This analysis was first performed in~\cite{Garcia:2023eol}. Characterizing the rapid redshift of the inflaton mass after fragmentation, $m_{\phi}\propto (a/a_{\rm end})^{1.32}$ from lattice data, the number density of free inflatons as given in Eq.~(\ref{eq:ndeltaphi}), and the decay rate
\beq\label{eq:gammadeltaphi}
\Gamma_{\delta\phi} \;=\; \frac{y^2}{8\pi}\,m_{\phi}\,,
\eeq
(neglecting the masses of the outgoing states), the authors obtain the following estimate for the reheating temperature for post-fragmentation decays:
\beq\label{eq:fragTreh}
T_{\rm reh} \;\simeq\; \left(\frac{30 \rho_{\rm end}}{\pi^2 g_{\rm reh}}\right)^{1/4} \left[ \frac{\gamma_{\phi}y^2}{8\pi(1-x)}\left(\frac{M_P^4}{\rho_{\rm end}}\right)^{3/2}\right]^{\frac{1}{1-x}}\,.
\eeq
Here $\gamma_{\phi}\simeq 2.49\times 10^{-15}$ and $x\simeq 0.65$ are lattice-derived constants. The main conclusion to be extracted from this result is that, under the aforementioned approximations, reheating to fermions in the quartic potential is impossible above $T_{\rm BBN}\simeq 4\,{\rm MeV}$ unless $y\gtrsim 2.7\times 10^{-4}$ ($y^2/\lambda\gtrsim 2\times 10^4$).

In the present work we have found, however, that for $y\gtrsim 10^{-7}$ resonance and quantum statistics effects play a major role in determining the efficiency of the reheating process prior to fragmentation. In particular, the Pauli bound $f_{\psi}=1$ will already be saturated for a wide range of momenta when fragmentation occurs. This implies that even after fragmentation we cannot make any approximations to determine the fermion production rate. Treating the post-fragmentation particle production process as a free-inflaton decay process, we must then solve the Boltzmann equation {\em in full}, with initial conditions determined by the distributions obtained in Section~\ref{sec:nonpert}.\par\medskip

The collision term which determines the evolution of the fermion PSD $f_{\psi}(P)$ from the decay $\delta\phi\rightarrow\bar{\psi}\psi$ is given by
\begin{align} \notag
    \mathcal{C}[f_{\psi}] 
    \;=\; \frac{1}{P^0} \int &\frac{\diff^3 \bK}{(2\pi)^3 2K^0} 
    \frac{\diff^3 \bP'}{(2\pi)^3 2P'^0} (2\pi)^4 \delta^{(4)}(K-P-P') 
    \abs{\overline{\mathcal{M}_{\delta\phi \to \psi\psi}}}^2\\ 
    &\times \left[f_{\delta\phi}(K) \left( 1 - f_\psi(P) - f_\psi(P') \right) - f_\psi(P) f_\psi(P')\right]\,.
\end{align}
Using the relation
\begin{equation}
    \abs{\overline{\mathcal{M}_{\delta\phi \to \psi\psi}}}^2 
    = 16\pi \beta^2 m_\phi  \,\Gamma_{\delta\phi}\,,\qquad \beta \;=\;   \sqrt{1-\dfrac{4 m_{\psi,{\rm eff}}^2(t)}{m_\phi^2(t)} }  \,,
\end{equation}
where $\Gamma_{\delta\phi}$ is given by (\ref{eq:gammadeltaphi}), and integrating with respect to $\bP'$ in the first term, and by $\bK$ in the second term, we obtain the following expansion 
\begin{align} \notag
    \mathcal{C}[f_{\psi}] 
    \;=\; \frac{2\beta^2\Gamma_{\delta\phi}m_{\phi}}{|\bP| P^0} \Bigg\{ &\int \frac{|\bK|\,\diff |\bK|}{K^0} 
    f_{\delta\phi}(K) \left[ 1 - f_\psi(P) - f_\psi \left( \sqrt{(K^0-P^0)^2-m_{\phi}^2} \right) \right] \theta\left( 1 - \left| \frac{2K^0P^0-m_{\phi}}{2|\bK||\bP|} \right| \right)\\ \label{eq:Boltzfull}
    & - f_{\psi}(P)\int \frac{|\bP'|\diff \bP'}{P^{\prime 0}}f_{\psi}(P')\ \theta\left( 1 - \left| \frac{2m_{\psi}^2-m_{\phi}^2+2P^{\prime0}P^0}{2|\bP'||\bP|} \right|\right)\Bigg\}\,,
\end{align}
where momenta are in the mass shell, e.g.~$P^0=\sqrt{\bP^2+m_{\psi, {\rm eff}}^2}$. The Boltzmann PDE can be converted into an ordinary (integro-)differential equation by switching to comoving momentum. The equation that we solve is then given by
\beq\label{eq:boltzfa}
\frac{\diff}{\diff t}f_{\psi}(p,t) \;=\; \langle \mathcal{C}[f_{\psi}]_{\bP\rightarrow \bp/a} \rangle\,,
\eeq
where we indicate in shorthand notation the replacement of all physical momenta ($K,P,P'$) in (\ref{eq:Boltzfull}) by comoving momenta. The brackets indicate averaging over the oscillation of the inflaton over each period. The inputs correspond to the initial condition for the PSD, which we obtain from the non-perturbative computation, and the inflaton PSD, which we extract from the lattice runs. The numerical implementation of Eq.~(\ref{eq:boltzfa}) is technically challenging,\footnote{We solve it by means of a custom Mathematica code.} and for that reason we limit out discussion to a small range of couplings, without weakening our final conclusions.

\begin{figure}
\centering
    \includegraphics[width=\textwidth]{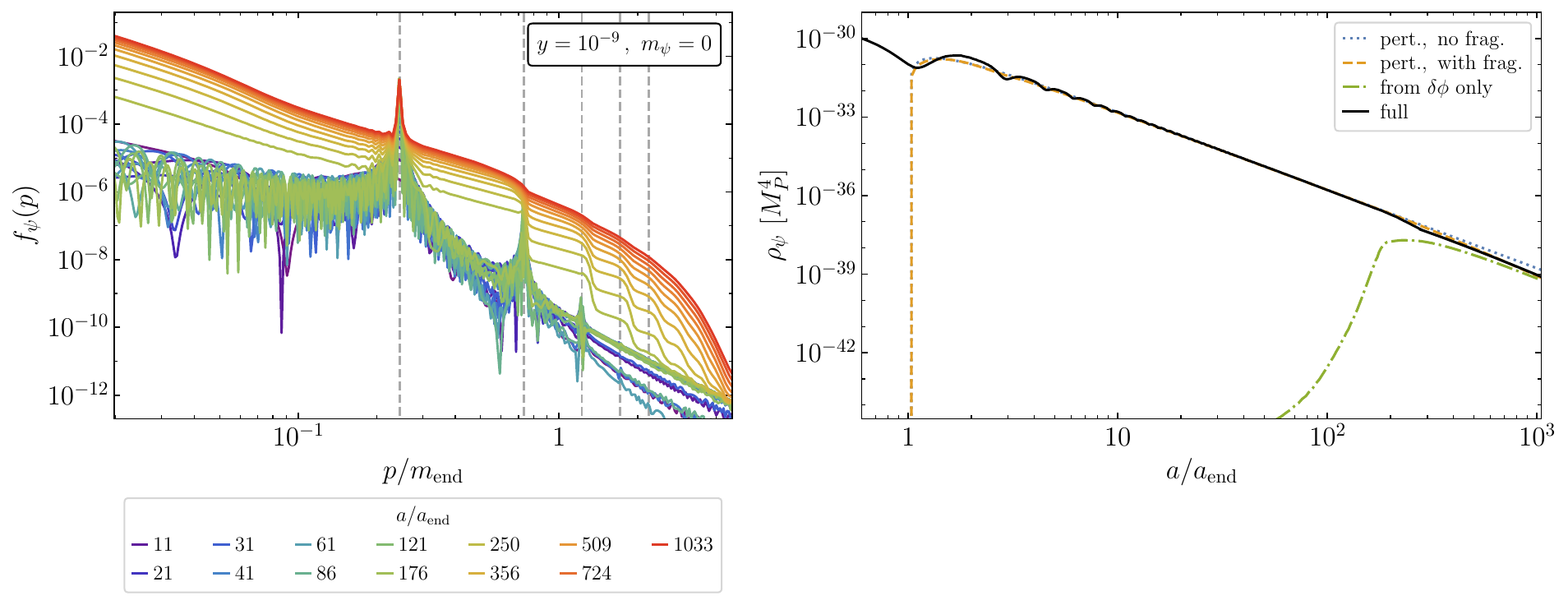}
     \caption{Left: Fermion PSD evolution before and after inflaton fragmentation for a Yukawa coupling $y=10^{-9}$. For $a/a_{\rm end}\leq 180$ the PSD is obtained from (\ref{eq:np1}), while for $a/a_{\rm end}>180$ it corresponds to the solution of (\ref{eq:boltzfa}). The time evolution is coded by color, with the caption showing the corresponding scale factors for every other curve. Vertical dashed lines show the first five Boltzmann peaks (see Fig.~\ref{fig:1e-8}). Right: Comparison of different approximations for the fermion energy density before and after inflaton fragmentation. The blue dotted curve shows the perturbative approximation (with $\mathcal R=0$) ignoring $\phi$-fragmentation. The orange dashed curve corresponds to the perturbative approximation (also with $\mathcal{R}=0$) accounting for fragmentation, by solving the system (\ref{eq:contfrag1})-(\ref{eq:contfrag2}). The dot-dashed green curve shows the energy density of $\psi$ assuming only $\delta\phi$ decays; that is, the solution of the full Boltzmann equation (\ref{eq:boltzfa}) with zero initial condition. The continuous black curve corresponds to the full solution for $\rho_{\psi}$, obtained from the integration of the exact PSD, shown in the left panel.}
    \label{fig:1e-9_full_energy_density}
\end{figure}

Fig.~\ref{fig:1e-9_full_energy_density} shows the result of integrating the full Boltzmann equation for a Yukawa coupling $y=10^{-9}$ and a vanishing bare mass of the fermion. The left panel demonstrates the time evolution of the PSD with initial conditions obtained from the solution of the Dirac equation until the onset of fragmentation at $a/a_{\rm end}=180$. After this point, we neglect the residual production of fermions from the left-over condensate, assuming that inflaton quanta decays take over the $\psi$ production. We note in this case that the widening of the inflaton PSD due to rescatterings leaves a clear imprint in the fermion PSD. The distribution ceases to be concentrated at discrete peaks, and is instead smoothly filled for all momenta for $p/m_{\rm end}\lesssim 5$; for larger momenta the production is exponentially suppressed. Only the main Boltzmann peak remains visible in the range of scale factors considered in the figure. As expected, for such a low coupling the PSD is far from saturating the Pauli limit, and we would expect our full computation to match the simple Boltzmann estimates. This is verified in the right panel of Fig.~\ref{fig:1e-9_full_energy_density}, where we show four different approximations to the computation of $\rho_{\psi}$. The first one (blue, dotted) corresponds to the Boltzmann approximation under the assumption of no inflaton fragmentation. The second one (yellow, dashed) shows the result of accounting for fragmentation neglecting Fermi-Dirac statistics and kinematic blocking (\ref{eq:contfrag1})-(\ref{eq:contfrag2}). The third one (green, dot-dashed) is the exact solution to the full Boltzmann equation (\ref{eq:boltzfa}) accounting only for $\delta\phi$ decays. And finally, the fourth one (black, continuous) represents the full solution for $\rho_{\psi}$, from non-perturbative methods prior to fragmentation, and the full solution to (\ref{eq:boltzfa}) after fragmentation. We note that all curves that account for the dissipation of the condensate $\phi$ agree with each other for $a/a_{\rm end}\lesssim 180$. After this point, fragmentation effects are important, but Pauli suppression is not, so the blue, dotted curve ceases to be correct, while the remaining three approximations agree after a transient period. Importantly we see explicitly that $\psi$ production is now driven by $\delta\phi$ decays. 

\begin{figure}
\centering
     \includegraphics[width=\textwidth]{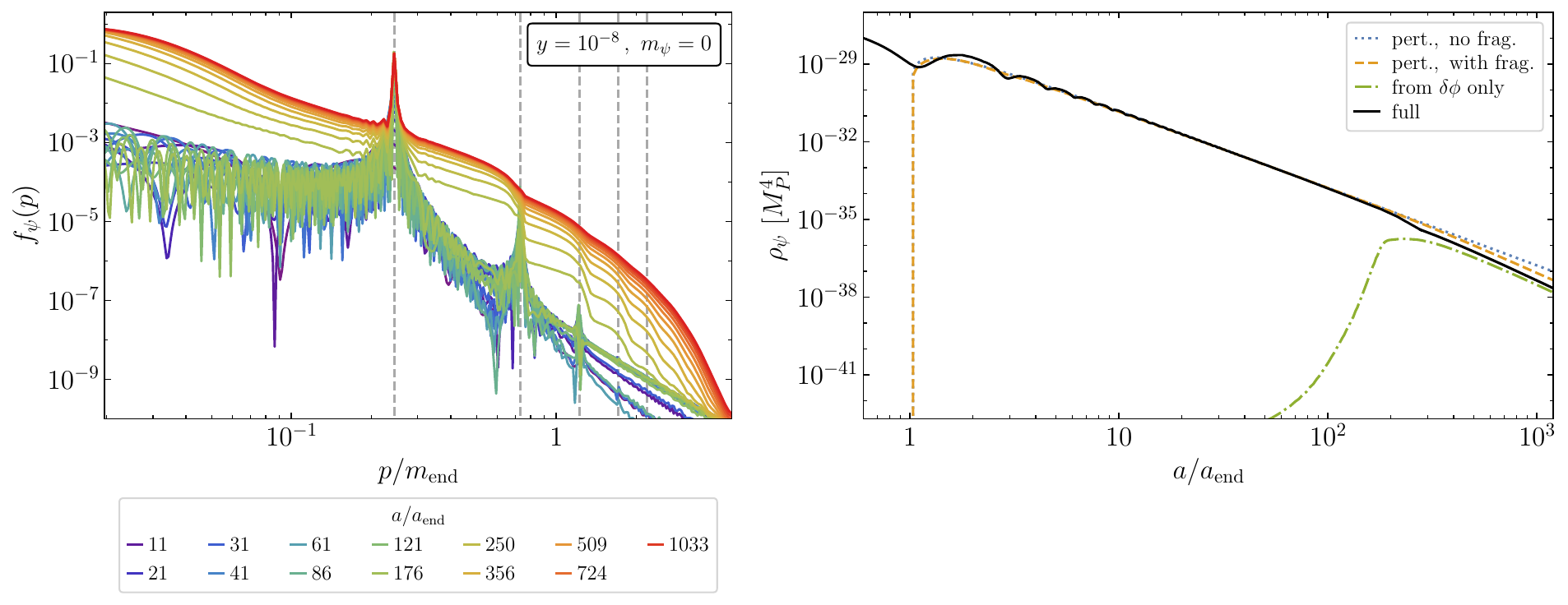}
     
            \caption{As in Fig.~\ref{fig:1e-9_full_energy_density}, the fermion PSD before and after fragmentation, for a Yukawa coupling $y=10^{-8}$ (left), and a comparison of different approximations to $\rho_{\psi}$ (right).}
    \label{fig:1e-8_full_energy_density}
\end{figure}

Fig.~\ref{fig:1e-8_full_energy_density} shows the evolution of the fermion sector for a larger coupling, $y=10^{-8}$ ($y^2/\lambda\simeq 3\times 10^{-5}$), still within the Boltzmann-validity range prior to fragmentation. Indeed, we observe that the initial PSD is always $f_{\psi}\ll 1$, except for the main peak, and that the exact energy density agrees with the perturbative approximation initially. Nevertheless, the resonant production of inflaton quanta leads to very high inflaton occupation numbers, $f_{\delta\phi}\lesssim 10^{13}$, and their decay even at this low value of $y$ leads to a rapid saturation of the fermion PSD. We therefore observe, in the right panel, that the exact energy density of $\psi$ is reduced relative to the approximation ignoring quantum statistics. Extrapolation of these results would imply an ever lower reheating temperature than that given by the approximation (\ref{eq:fragTreh}).

\begin{figure}
\centering
     \includegraphics[width=\textwidth]{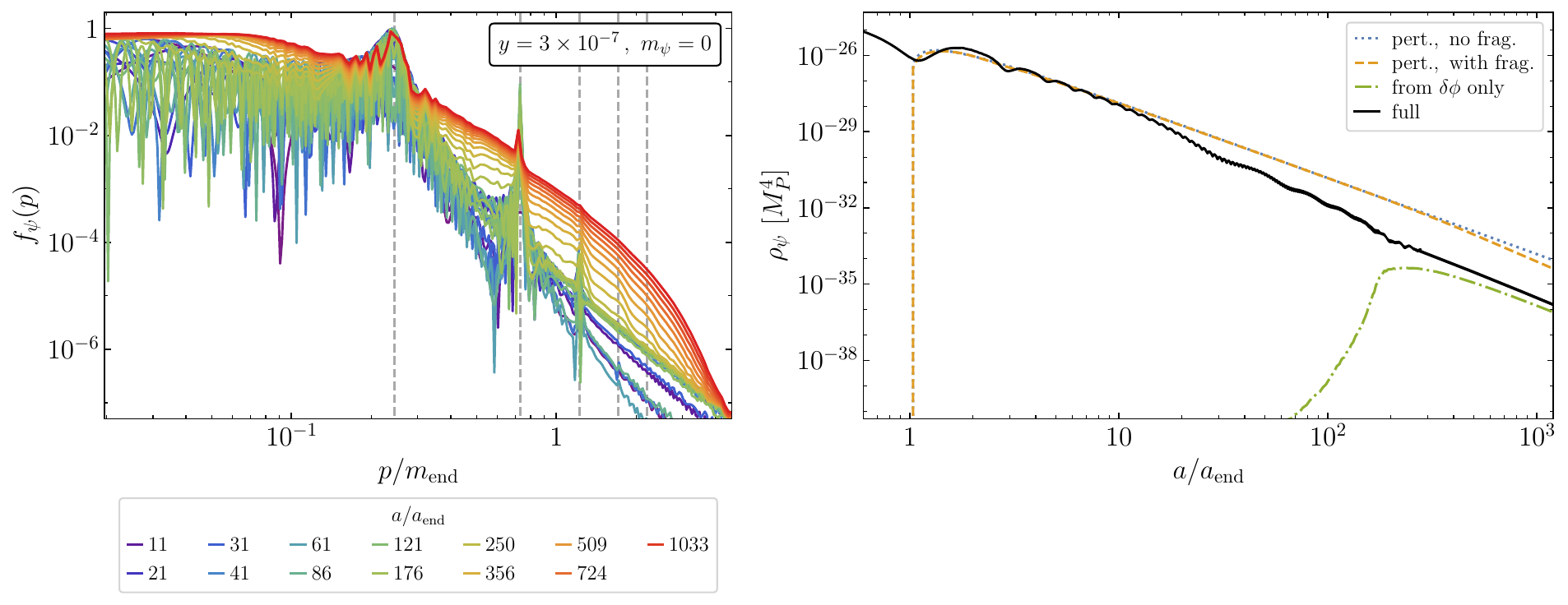}
     
            \caption{As in Fig.~\ref{fig:1e-9_full_energy_density}, the fermion PSD before and after fragmentation, for a Yukawa coupling $y=3\times 10^{-7}$ (left), and a comparison of different approximations to $\rho_{\psi}$ (right).}
    \label{fig:3e-7_full_energy_density}
\end{figure}

In Fig.~\ref{fig:3e-7_full_energy_density} we show the result of our exact post-fragmentation computation for a coupling slightly above the pre-fragmentation non-perturbative threshold, $y=3\times 10^{-7}$ ($y^2/\lambda\simeq 3\times 10^{-2}$). As it is clear from the left panel, with an initial $f_{\psi}\sim 1$ at low momenta, we would not only expect the simple perturbative approximation to fail shortly after the beginning of reheating, but we would also expect that accounting only for $\delta\phi$ decays after fragmentation would also fail to correctly account for the amount of produced fermions, as the PSD comes ``pre-saturated''. We indeed observe (in the right panel) a clear deviation of the exact solution relative to the perturbative approximation starting at $a/a_{\rm end}\simeq 10$. After fragmentation, this separation persists, and the estimates in the energy density differ by more than one order of magnitude. Similarly, the presence of the non-perturbative peak at $p/m_{\rm end}\simeq 0.25$ implies that the dot-dashed line underestimates the correct value of $\rho_{\psi}$. 

The solution of the full Boltzmann equation (\ref{eq:boltzfa}) becomes increasingly computationally challenging for larger couplings, and for that reason we do not show the corresponding results. Nevertheless, the trend is clear: for larger couplings the Pauli suppression will be stronger, and therefore (\ref{eq:fragTreh}) will over estimate the correct reheating temperature by several orders of magnitude. Therefore, unless alternative (bosonic) decay channels for $\phi$ exist, or thermalization helps alleviate the saturation of fermionic levels, reheating will not be completed above the BBN temperature if it does not occur prior to fragmentation.


\section{Conclusions}\label{sec:conclusion}

In the present work, we have considered the decay of the inflaton field into spin 1/2 fermions when the inflationary potential near the minima can be approximated by a quartic term. The main idea behind this study is to investigate the interplay between fermion particle production and inflaton fragmentation. While we focus on the quartic T-model potential for concreteness, our conclusions should remain largely insensitive to the form of the potential far from the minimum in large-field inflation scenarios. Models with quartic minima have the peculiarity that the number of $e$-folds are independent on the details of the reheating mechanism, since radiation dominates immediately after the onset of oscillations. The sole constraint on reheating corresponds to the lower bound on the reheating temperature, which is to be above the primordial nucleosynthesis scale. However, additional restrictions on $T_{\rm reh}$ may arise from different phenomenological considerations, such as dark matter production or baryogenesis.

For the previous reasons, it is tempting to assume that the decay of the inflaton in the quartic scenario occurs perturbatively in the Boltzmann sense: all the sensitivity of the dynamics to the short time-scale oscillations of the inflaton can be averaged away, with the decay rate determining the efficiency of particle production and the reheating temperature. For an inflaton that decays into bosonic degrees of freedom, this is not a generic possibility, as the kinematic parameter that determines the deviation from adiabaticity of the system, $\mathcal{R}$, increases with time~\cite{Garcia:2020wiy}. Therefore, in many instances fermionic final states for inflaton decays are invoked to avoid the presence of parametric resonance and bosonic enhancement effects in the estimation of the duration of the reheating epoch. However, we find that in the case of quartic potentials, non perturbative effects cannot be avoided for fermionic decay products. For the range of couplings required for successful reheating~\cite{Garcia:2023eol}, the presence of resonance effects, kinematic blocking and Pauli suppression halt the decays of the inflaton, such that, in the absence of no other coupling to matter, reheating is not possible.

In order to obtain our results we have taken a series of assumptions. Firstly, throughout our
work we have assumed that the fermionic decay products are much lighter than the inflaton at all times, simulating them as strictly massless (in the bare sense). We have neglected the presence of a bare inflaton mass, which might take over the dynamics of reheating at late times. For instance, its presence would alleviate the suppression of the inflaton decay rate, and allow the completion of reheating for couplings as small as $y\gtrsim 2\times 10^{-11}(m_{\phi}/1\,{\rm GeV})^{-1/2}$~\cite{Clery:2024dlk}.
Furthermore, since we have not considered any explicit coupling of $\psi$ to other light particles, we also do not consider thermal effects. Lastly, in this work we limited ourselves to the study of fermion production, regardless of its origin. For this purpose, we have also considered the large Yukawa coupling regime, where loop corrections play an important role. Our results in this regime should be interpreted as a rather conservative bound, as we specifically restrain ourselves from quantifying the backreaction effects on the inflaton dynamics.
We can summarize our findings as follows:
\begin{itemize}
\item \textbf{For $y,y_5\gtrsim 10^{-7}$ ($y^2/\lambda\gtrsim 3\times 10^{-3}$):}\\
This is the range that contains the phenomenologically viable scenarios. We find that, the Boltzmann equation fails to provide an adequate approximation to the dynamics in this regime. We can differentiate two regimes: post- and pre-fragmentation. If reheating is not complete before fragmentation, the initial conditions for post-fragmentation must be obtained from the solution of the Heisenberg equations of motion.  If on the other hand, reheating is complete before fragmentation, a full account of fermionic backreaction is needed to obtain an estimate of the reheating temperature. 
\item \textbf{For couplings $y\gtrsim 10^{-8}$ ($y^2/\lambda\gtrsim 3\times 10^{-5}$)}:\\
In this case, for the post-fragmentation decays of inflaton quanta, we make use of the full Boltzmann equation to show that Pauli suppression is present, leading to a much less efficient reheating than what would be obtained ignoring quantum statistics.
\end{itemize}

We conclude that fermion reheating in a quartic potential relies necessarily on large couplings, i.e, $y\gtrsim 0.2$ ($y^2/\lambda\simeq 10^{10}$), or on a rapid thermalization of $\psi$ with Standard Model degrees of freedom that alleviates the saturation of the fermion energy levels. The time-scale for such thermalization is however unknown. It is worth nothing that our results are mostly applicable for the production of dark sector fermions, if reheating is completed by alternative decay channels. We will explore in detail this possibility in later work, as the aforementioned alternative inflaton decay channels will likely require a non-perturbative treatment as well.

Since we assumed the fermionic decay products to be light, all kinematic effects arise from their Yukawa-like coupling to the inflaton. It is of interest to explore the large mass regime, to determine how the success of reheating is modified for Standard Model-adjacent $\psi$, or in the case of heavy dark matter $\psi$. Additionally, it is worth exploring fermion reheating in detail in potentials of the form $V(\phi)\propto\phi^k$ with $k>4$. In the perturbative approximation, it is known that fermion reheating after fragmentation is impossible~\cite{Garcia:2023dyf}, but little is known about the relevance of resonance/suppression effects for pre-fragmentation particle production.


\section*{Acknowledgments}
 EC and MG are supported by the DGAPA-PAPIIT grant IA100525 at UNAM, the SECIHTI “Ciencia de Frontera” grant CF-2023-I-17, and a Cátedra Marcos Moshinsky. NB, AM and MP acknowledge support by the Deutsche Forschungsgemeinschaft (DFG, German Research Foundation) under the DFG Emmy Noether Grant No. PI 1933/1-1 and Germany's Excellence Strategy – EXC 2121 “Quantum Universe” – 390833306. This research was supported in part through the Maxwell computational resources operated at Deutsches Elektronen-Synchrotron DESY, Hamburg,
Germany. NB and MG acknowledge support by Institut Pascal and the P2I axis of the Graduate School of Physics during the Paris-Saclay Astroparticle Symposium 2025, as well as the CNRS IRP UCMN.


\addcontentsline{toc}{section}{References}
\bibliographystyle{utphys}
\bibliography{references} 

@book{Linde:1990flp,
    author = "Linde, Andrei D.",
    title = "{Particle physics and inflationary cosmology}",
    eprint = "hep-th/0503203",
    archivePrefix = "arXiv",
    volume = "5",
    year = "1990"
}

@article{Lyth:1998xn,
    author = "Lyth, David H. and Riotto, Antonio",
    title = "{Particle physics models of inflation and the cosmological density perturbation}",
    eprint = "hep-ph/9807278",
    archivePrefix = "arXiv",
    reportNumber = "LANCS-TH-9720, FERMILAB-PUB-97-292-A, CERN-TH-97-383, OUTP-98-39-P",
    doi = "10.1016/S0370-1573(98)00128-8",
    journal = "Phys. Rept.",
    volume = "314",
    pages = "1--146",
    year = "1999"
}

@article{Linde:2000kn,
    author = "Linde, Andrei D.",
    title = "{Inflationary cosmology}",
    doi = "10.1016/S0370-1573(00)00038-7",
    journal = "Phys. Rept.",
    volume = "333",
    pages = "575--591",
    year = "2000"
}

@article{Starobinsky:1980te,
    author = "Starobinsky, Alexei A.",
    editor = "Khalatnikov, I. M. and Mineev, V. P.",
    title = "{A New Type of Isotropic Cosmological Models Without Singularity}",
    doi = "10.1016/0370-2693(80)90670-X",
    journal = "Phys. Lett. B",
    volume = "91",
    pages = "99--102",
    year = "1980"
}

@article{Kallosh:2013hoa,
    author = "Kallosh, Renata and Linde, Andrei",
    title = "{Universality Class in Conformal Inflation}",
    eprint = "1306.5220",
    archivePrefix = "arXiv",
    primaryClass = "hep-th",
    doi = "10.1088/1475-7516/2013/07/002",
    journal = "JCAP",
    volume = "07",
    pages = "002",
    year = "2013"
}

@article{Planck:2018vyg,
    author = "Aghanim, N. and others",
    collaboration = "Planck",
    title = "{Planck 2018 results. VI. Cosmological parameters}",
    eprint = "1807.06209",
    archivePrefix = "arXiv",
    primaryClass = "astro-ph.CO",
    doi = "10.1051/0004-6361/201833910",
    journal = "Astron. Astrophys.",
    volume = "641",
    pages = "A6",
    year = "2020",
    note = "[Erratum: Astron.Astrophys. 652, C4 (2021)]"
}

@article{Planck:2018jri,
    author = "Akrami, Y. and others",
    collaboration = "Planck",
    title = "{Planck 2018 results. X. Constraints on inflation}",
    eprint = "1807.06211",
    archivePrefix = "arXiv",
    primaryClass = "astro-ph.CO",
    doi = "10.1051/0004-6361/201833887",
    journal = "Astron. Astrophys.",
    volume = "641",
    pages = "A10",
    year = "2020"
}

@article{BICEP:2021xfz,
    author = "Ade, P. A. R. and others",
    collaboration = "BICEP, Keck",
    title = "{Improved Constraints on Primordial Gravitational Waves using Planck, WMAP, and BICEP/Keck Observations through the 2018 Observing Season}",
    eprint = "2110.00483",
    archivePrefix = "arXiv",
    primaryClass = "astro-ph.CO",
    doi = "10.1103/PhysRevLett.127.151301",
    journal = "Phys. Rev. Lett.",
    volume = "127",
    number = "15",
    pages = "151301",
    year = "2021"
}

@article{Tristram:2021tvh,
    author = "Tristram, M. and others",
    title = "{Improved limits on the tensor-to-scalar ratio using BICEP and Planck data}",
    eprint = "2112.07961",
    archivePrefix = "arXiv",
    primaryClass = "astro-ph.CO",
    doi = "10.1103/PhysRevD.105.083524",
    journal = "Phys. Rev. D",
    volume = "105",
    number = "8",
    pages = "083524",
    year = "2022"
}

@article{ACT:2025fju,
    author = "Louis, Thibaut and others",
    collaboration = "ACT",
    title = "{The Atacama Cosmology Telescope: DR6 Power Spectra, Likelihoods and $\Lambda$CDM Parameters}",
    eprint = "2503.14452",
    archivePrefix = "arXiv",
    primaryClass = "astro-ph.CO",
    reportNumber = "FERMILAB-PUB-25-0071-PPD",
    month = "3",
    year = "2025"
}

@article{ACT:2025tim,
    author = "Calabrese, Erminia and others",
    collaboration = "ACT",
    title = "{The Atacama Cosmology Telescope: DR6 Constraints on Extended Cosmological Models}",
    eprint = "2503.14454",
    archivePrefix = "arXiv",
    primaryClass = "astro-ph.CO",
    reportNumber = "FERMILAB-PUB-25-0157-PPD",
    month = "3",
    year = "2025"
}

@article{SPT-3G:2025bzu,
    author = "Camphuis, E. and others",
    collaboration = "SPT-3G",
    title = "{SPT-3G D1: CMB temperature and polarization power spectra and cosmology from 2019 and 2020 observations of the SPT-3G Main field}",
    eprint = "2506.20707",
    archivePrefix = "arXiv",
    primaryClass = "astro-ph.CO",
    reportNumber = "FERMILAB-PUB-25-0144-PPD",
    month = "6",
    year = "2025"
}

@article{Ellis:2025bzi,
    author = "Ellis, John and Gherghetta, Tony and Kaneta, Kunio and Ke, Wenqi and Olive, Keith A.",
    title = "{Effects of radiative corrections on Starobinsky inflation}",
    eprint = "2510.15137",
    archivePrefix = "arXiv",
    primaryClass = "hep-ph",
    reportNumber = "UMN-TH-4511/25, FTPI-MINN-25/13, KCL-PH-TH/2025-39, CERN-TH-2025-198",
    doi = "10.1103/8cx9-c642",
    journal = "Phys. Rev. D",
    volume = "112",
    number = "12",
    pages = "123530",
    year = "2025"
}

@article{Alexandre:2025ixz,
    author = "Alexandre, Jean and Heurtier, Lucien and Pla, Silvia",
    title = "{Exact Renormalisation Group Evolution of the Inflation Dynamics: Reconciling $\alpha$-Attractors with ACT}",
    eprint = "2511.05296",
    archivePrefix = "arXiv",
    primaryClass = "hep-th",
    reportNumber = "KCL-PH-TH/2025-46, TUM-HEP-1578/25",
    month = "11",
    year = "2025"
}

@article{Gross:2015bea,
    author = "Gross, Christian and Lebedev, Oleg and Zatta, Marco",
    title = "{Higgs{\textendash}inflaton coupling from reheating and the metastable Universe}",
    eprint = "1506.05106",
    archivePrefix = "arXiv",
    primaryClass = "hep-ph",
    reportNumber = "HIP-2015-39-TH",
    doi = "10.1016/j.physletb.2015.12.014",
    journal = "Phys. Lett. B",
    volume = "753",
    pages = "178--181",
    year = "2016"
}

@article{Garcia:2020wiy,
    author = "Garcia, Marcos A. G. and Kaneta, Kunio and Mambrini, Yann and Olive, Keith A.",
    title = "{Inflaton Oscillations and Post-Inflationary Reheating}",
    eprint = "2012.10756",
    archivePrefix = "arXiv",
    primaryClass = "hep-ph",
    reportNumber = "UMN-TH-4006/20, FTPI-MINN-20/37, IFT-UAM/CSIC-20-185, KIAS-P20071",
    doi = "10.1088/1475-7516/2021/04/012",
    journal = "JCAP",
    volume = "04",
    pages = "012",
    year = "2021"
}

@article{Ballesteros:2020adh,
    author = "Ballesteros, Guillermo and Garcia, Marcos A. G. and Pierre, Mathias",
    title = "{How warm are non-thermal relics? Lyman-$\alpha$ bounds on out-of-equilibrium dark matter}",
    eprint = "2011.13458",
    archivePrefix = "arXiv",
    primaryClass = "hep-ph",
    reportNumber = "IFT-UAM/CSIC-20-135",
    doi = "10.1088/1475-7516/2021/03/101",
    journal = "JCAP",
    volume = "03",
    pages = "101",
    year = "2021"
}

@article{Kofman:1997yn,
    author = "Kofman, Lev and Linde, Andrei D. and Starobinsky, Alexei A.",
    title = "{Towards the theory of reheating after inflation}",
    eprint = "hep-ph/9704452",
    archivePrefix = "arXiv",
    reportNumber = "IFA-97-28, SU-ITP-97-18",
    doi = "10.1103/PhysRevD.56.3258",
    journal = "Phys. Rev. D",
    volume = "56",
    pages = "3258--3295",
    year = "1997"
}

@article{Nurmi:2015ema,
    author = "Nurmi, Sami and Tenkanen, Tommi and Tuominen, Kimmo",
    title = "{Inflationary Imprints on Dark Matter}",
    eprint = "1506.04048",
    archivePrefix = "arXiv",
    primaryClass = "astro-ph.CO",
    reportNumber = "HIP-2015-6-TH",
    doi = "10.1088/1475-7516/2015/11/001",
    journal = "JCAP",
    volume = "11",
    pages = "001",
    year = "2015"
}

@article{Martin:2010kz,
    author = "Martin, Jerome and Ringeval, Christophe",
    title = "{First CMB Constraints on the Inflationary Reheating Temperature}",
    eprint = "1004.5525",
    archivePrefix = "arXiv",
    primaryClass = "astro-ph.CO",
    doi = "10.1103/PhysRevD.82.023511",
    journal = "Phys. Rev. D",
    volume = "82",
    pages = "023511",
    year = "2010"
}

@article{Garcia:2020eof,
    author = "Garcia, Marcos A. G. and Kaneta, Kunio and Mambrini, Yann and Olive, Keith A.",
    title = "{Reheating and Post-inflationary Production of Dark Matter}",
    eprint = "2004.08404",
    archivePrefix = "arXiv",
    primaryClass = "hep-ph",
    reportNumber = "UMN--TH--3916/20, FTPI--MINN--20/06, IFT-UAM/CSIC-20-56",
    doi = "10.1103/PhysRevD.101.123507",
    journal = "Phys. Rev. D",
    volume = "101",
    number = "12",
    pages = "123507",
    year = "2020"
}

@article{Turner:1983he,
    author = "Turner, Michael S.",
    title = "{Coherent Scalar Field Oscillations in an Expanding Universe}",
    reportNumber = "EFI-83-29-CHICAGO",
    doi = "10.1103/PhysRevD.28.1243",
    journal = "Phys. Rev. D",
    volume = "28",
    pages = "1243",
    year = "1983"
}

@article{Ichikawa:2008ne,
    author = "Ichikawa, Kazuhide and Suyama, Teruaki and Takahashi, Tomo and Yamaguchi, Masahide",
    title = "{Primordial Curvature Fluctuation and Its Non-Gaussianity in Models with Modulated Reheating}",
    eprint = "0807.3988",
    archivePrefix = "arXiv",
    primaryClass = "astro-ph",
    doi = "10.1103/PhysRevD.78.063545",
    journal = "Phys. Rev. D",
    volume = "78",
    pages = "063545",
    year = "2008"
}

@article{Boyle:2018rgh,
    author = "Boyle, Latham and Finn, Kieran and Turok, Neil",
    title = "{The Big Bang, CPT, and neutrino dark matter}",
    eprint = "1803.08930",
    archivePrefix = "arXiv",
    primaryClass = "hep-ph",
    doi = "10.1016/j.aop.2022.168767",
    journal = "Annals Phys.",
    volume = "438",
    pages = "168767",
    year = "2022"
}

@article{Kolb:2023ydq,
    author = "Kolb, Edward W. and Long, Andrew J.",
    title = "{Cosmological gravitational particle production and its implications for cosmological relics}",
    eprint = "2312.09042",
    archivePrefix = "arXiv",
    primaryClass = "astro-ph.CO",
    doi = "10.1103/RevModPhys.96.045005",
    journal = "Rev. Mod. Phys.",
    volume = "96",
    number = "4",
    pages = "045005",
    year = "2024"
}

@article{Kuzmin:1998kk,
    author = "Kuzmin, Vadim and Tkachev, Igor",
    title = "{Matter creation via vacuum fluctuations in the early universe and observed ultrahigh-energy cosmic ray events}",
    eprint = "hep-ph/9809547",
    archivePrefix = "arXiv",
    reportNumber = "PURD-TH-98-10",
    doi = "10.1103/PhysRevD.59.123006",
    journal = "Phys. Rev. D",
    volume = "59",
    pages = "123006",
    year = "1999"
}

@article{Chung:2011ck,
    author = "Chung, Daniel J. H. and Everett, Lisa L. and Yoo, Hojin and Zhou, Peng",
    title = "{Gravitational Fermion Production in Inflationary Cosmology}",
    eprint = "1109.2524",
    archivePrefix = "arXiv",
    primaryClass = "astro-ph.CO",
    doi = "10.1016/j.physletb.2012.04.066",
    journal = "Phys. Lett. B",
    volume = "712",
    pages = "147--154",
    year = "2012"
}

@article{Garcia:2021iag,
    author = "Garcia, Marcos A. G. and Kaneta, Kunio and Mambrini, Yann and Olive, Keith A. and Verner, Sarunas",
    title = "{Freeze-in from preheating}",
    eprint = "2109.13280",
    archivePrefix = "arXiv",
    primaryClass = "hep-ph",
    reportNumber = "UMN-TH-4101/21, FTPI-MINN-21/19, CERN-TH-2021-121",
    doi = "10.1088/1475-7516/2022/03/016",
    journal = "JCAP",
    volume = "03",
    number = "03",
    pages = "016",
    year = "2022"
}

@article{Figueroa:2020rrl,
    author = "Figueroa, Daniel G. and Florio, Adrien and Torrenti, Francisco and Valkenburg, Wessel",
    title = "{The art of simulating the early Universe -- Part I}",
    eprint = "2006.15122",
    archivePrefix = "arXiv",
    primaryClass = "astro-ph.CO",
    doi = "10.1088/1475-7516/2021/04/035",
    journal = "JCAP",
    volume = "04",
    pages = "035",
    year = "2021"
}

@article{Figueroa:2021yhd,
    author = "Figueroa, Daniel G. and Florio, Adrien and Torrenti, Francisco and Valkenburg, Wessel",
    title = "{CosmoLattice}",
    eprint = "2102.01031",
    archivePrefix = "arXiv",
    primaryClass = "astro-ph.CO",
    month = "2",
    year = "2021"
}

@article{Amin:2011hj,
    author = "Amin, Mustafa A. and Easther, Richard and Finkel, Hal and Flauger, Raphael and Hertzberg, Mark P.",
    title = "{Oscillons After Inflation}",
    eprint = "1106.3335",
    archivePrefix = "arXiv",
    primaryClass = "astro-ph.CO",
    doi = "10.1103/PhysRevLett.108.241302",
    journal = "Phys. Rev. Lett.",
    volume = "108",
    pages = "241302",
    year = "2012"
}

@article{Lebedev:2022vwf,
    author = "Lebedev, Oleg and Solomko, Timofey and Yoon, Jong-Hyun",
    title = "{Dark matter production via a~non-minimal coupling to gravity}",
    eprint = "2211.11773",
    archivePrefix = "arXiv",
    primaryClass = "hep-ph",
    doi = "10.1088/1475-7516/2023/02/035",
    journal = "JCAP",
    volume = "02",
    pages = "035",
    year = "2023"
}

@article{Fu:2017ero,
    author = "Fu, Chengjie and Wu, Puxun and Yu, Hongwei",
    title = "{Production of gravitational waves during preheating with nonminimal coupling}",
    eprint = "1711.10888",
    archivePrefix = "arXiv",
    primaryClass = "gr-qc",
    doi = "10.1103/PhysRevD.97.081303",
    journal = "Phys. Rev. D",
    volume = "97",
    number = "8",
    pages = "081303",
    year = "2018"
}

@article{Figueroa:2016wxr,
    author = "Figueroa, Daniel G. and Torrenti, Francisco",
    title = "{Parametric resonance in the early Universe\textemdash{}a fitting analysis}",
    eprint = "1609.05197",
    archivePrefix = "arXiv",
    primaryClass = "astro-ph.CO",
    reportNumber = "CERN-TH-2016-202, IFT-UAM-CSIC-16-084",
    doi = "10.1088/1475-7516/2017/02/001",
    journal = "JCAP",
    volume = "02",
    pages = "001",
    year = "2017"
}

@article{Shtanov:1994ce,
    author = "Shtanov, Y. and Traschen, Jennie H. and Brandenberger, Robert H.",
    title = "{Universe reheating after inflation}",
    eprint = "hep-ph/9407247",
    archivePrefix = "arXiv",
    reportNumber = "BROWN-HET-957",
    doi = "10.1103/PhysRevD.51.5438",
    journal = "Phys. Rev. D",
    volume = "51",
    pages = "5438--5455",
    year = "1995"
}

@article{Lozanov:2016hid,
    author = "Lozanov, Kaloian D. and Amin, Mustafa A.",
    title = "{Equation of State and Duration to Radiation Domination after Inflation}",
    eprint = "1608.01213",
    archivePrefix = "arXiv",
    primaryClass = "astro-ph.CO",
    doi = "10.1103/PhysRevLett.119.061301",
    journal = "Phys. Rev. Lett.",
    volume = "119",
    number = "6",
    pages = "061301",
    year = "2017"
}

@article{Lozanov:2017hjm,
    author = "Lozanov, Kaloian D. and Amin, Mustafa A.",
    title = "{Self-resonance after inflation: oscillons, transients and radiation domination}",
    eprint = "1710.06851",
    archivePrefix = "arXiv",
    primaryClass = "astro-ph.CO",
    doi = "10.1103/PhysRevD.97.023533",
    journal = "Phys. Rev. D",
    volume = "97",
    number = "2",
    pages = "023533",
    year = "2018"
}

@article{Antusch:2021aiw,
    author = "Antusch, Stefan and Figueroa, Daniel G. and Marschall, Kenneth and Torrenti, Francisco",
    title = "{Characterizing the postinflationary reheating history: Single daughter field with quadratic-quadratic interaction}",
    eprint = "2112.11280",
    archivePrefix = "arXiv",
    primaryClass = "astro-ph.CO",
    doi = "10.1103/PhysRevD.105.043532",
    journal = "Phys. Rev. D",
    volume = "105",
    number = "4",
    pages = "043532",
    year = "2022"
}

@article{Amin:2014eta,
    author = "Amin, Mustafa A. and Hertzberg, Mark P. and Kaiser, David I. and Karouby, Johanna",
    title = "{Nonperturbative Dynamics Of Reheating After Inflation: A Review}",
    eprint = "1410.3808",
    archivePrefix = "arXiv",
    primaryClass = "hep-ph",
    doi = "10.1142/S0218271815300037",
    journal = "Int. J. Mod. Phys. D",
    volume = "24",
    pages = "1530003",
    year = "2014"
}

@BOOK{Magnus2004-br,
  title     = "Hill's Equation",
  author    = "Magnus, Wilhelm and Winkler, Stanley",
  publisher = "Dover Publications",
  series    = "Dover Books on Mathematics",
  month     =  feb,
  year      =  2004,
  address   = "Mineola, NY"
}

@article{Garcia:2022vwm,
    author = "Garcia, Marcos A. G. and Pierre, Mathias and Verner, Sarunas",
    title = "{Scalar dark matter production from preheating and structure formation constraints}",
    eprint = "2206.08940",
    archivePrefix = "arXiv",
    primaryClass = "hep-ph",
    reportNumber = "DESY-22-104",
    doi = "10.1103/PhysRevD.107.043530",
    journal = "Phys. Rev. D",
    volume = "107",
    number = "4",
    pages = "043530",
    year = "2023"
}

@article{Clery:2024dlk,
    author = "Clery, Simon and Garcia, Marcos A. G. and Mambrini, Yann and Olive, Keith A.",
    title = "{Bare mass effects on the reheating process after inflation}",
    eprint = "2402.16958",
    archivePrefix = "arXiv",
    primaryClass = "hep-ph",
    reportNumber = "UMN--TH--4325/24, FTPI--MINN--24/06",
    doi = "10.1103/PhysRevD.109.103540",
    journal = "Phys. Rev. D",
    volume = "109",
    number = "10",
    pages = "103540",
    year = "2024"
}

@article{Greene:1998nh,
    author = "Greene, Patrick B. and Kofman, Lev",
    title = "{Preheating of fermions}",
    eprint = "hep-ph/9807339",
    archivePrefix = "arXiv",
    reportNumber = "UH-IFA-98-44",
    doi = "10.1016/S0370-2693(99)00020-9",
    journal = "Phys. Lett. B",
    volume = "448",
    pages = "6--12",
    year = "1999"
}

@article{Garcia-Bellido:2000woy,
    author = "Garcia-Bellido, Juan and Mollerach, Silvia and Roulet, Esteban",
    title = "{Fermion production during preheating after hybrid inflation}",
    eprint = "hep-ph/0002076",
    archivePrefix = "arXiv",
    reportNumber = "FT-UAM-00-05, IFT-UAM-CSIC-00-05",
    doi = "10.1088/1126-6708/2000/02/034",
    journal = "JHEP",
    volume = "02",
    pages = "034",
    year = "2000"
}

@article{Giudice:1999fb,
    author = "Giudice, G. F. and Peloso, M. and Riotto, A. and Tkachev, I.",
    title = "{Production of massive fermions at preheating and leptogenesis}",
    eprint = "hep-ph/9905242",
    archivePrefix = "arXiv",
    reportNumber = "CERN-TH-99-117",
    doi = "10.1088/1126-6708/1999/08/014",
    journal = "JHEP",
    volume = "08",
    pages = "014",
    year = "1999"
}

@article{Greene:2000ew,
    author = "Greene, Patrick B. and Kofman, Lev",
    title = "{On the theory of fermionic preheating}",
    eprint = "hep-ph/0003018",
    archivePrefix = "arXiv",
    reportNumber = "CITA-2000-05",
    doi = "10.1103/PhysRevD.62.123516",
    journal = "Phys. Rev. D",
    volume = "62",
    pages = "123516",
    year = "2000"
}

@article{Berges:2010zv,
    author = "Berges, J. and Gelfand, D. and Pruschke, J.",
    title = "{Quantum theory of fermion production after inflation}",
    eprint = "1012.4632",
    archivePrefix = "arXiv",
    primaryClass = "hep-ph",
    doi = "10.1103/PhysRevLett.107.061301",
    journal = "Phys. Rev. Lett.",
    volume = "107",
    pages = "061301",
    year = "2011"
}

@article{Peloso:2000hy,
    author = "Peloso, Marco and Sorbo, Lorenzo",
    title = "{Preheating of massive fermions after inflation: Analytical results}",
    eprint = "hep-ph/0003045",
    archivePrefix = "arXiv",
    doi = "10.1088/1126-6708/2000/05/016",
    journal = "JHEP",
    volume = "05",
    pages = "016",
    year = "2000"
}

@article{Nilles:2001fg,
    author = "Nilles, Hans Peter and Peloso, Marco and Sorbo, Lorenzo",
    title = "{Coupled fields in external background with application to nonthermal production of gravitinos}",
    eprint = "hep-th/0103202",
    archivePrefix = "arXiv",
    doi = "10.1088/1126-6708/2001/04/004",
    journal = "JHEP",
    volume = "04",
    pages = "004",
    year = "2001"
}

@article{Olive:1989nu,
    author = "Olive, Keith A.",
    title = "{Inflation}",
    reportNumber = "UMN-TH-804-89",
    doi = "10.1016/0370-1573(90)90144-Q",
    journal = "Phys. Rept.",
    volume = "190",
    pages = "307--403",
    year = "1990"
}

@article{Dolgov:1989us,
    author = "Dolgov, A. D. and Kirilova, D. P.",
    title = "{On Particle Creation by a Time Dependent Scalar Field}",
    reportNumber = "JINR-E2-89-321",
    journal = "Sov. J. Nucl. Phys.",
    volume = "51",
    pages = "172--177",
    year = "1990"
}

@article{Traschen:1990sw,
    author = "Traschen, Jennie H. and Brandenberger, Robert H.",
    title = "{Particle Production During Out-of-equilibrium Phase Transitions}",
    reportNumber = "BROWN-HET-731",
    doi = "10.1103/PhysRevD.42.2491",
    journal = "Phys. Rev. D",
    volume = "42",
    pages = "2491--2504",
    year = "1990"
}

@article{Boyanovsky:1995ud,
    author = "Boyanovsky, D. and D'Attanasio, M. and de Vega, H. J. and Holman, R. and Lee, D. -S. and Singh, A.",
    title = "{Reheating the postinflationary universe}",
    eprint = "hep-ph/9505220",
    archivePrefix = "arXiv",
    reportNumber = "PITT-09-95, CMU-95-03, DOE-ER-40682-92, LPTHE-95-18, UPRF-95-420",
    month = "5",
    year = "1995"
}

@article{Yoshimura:1995gc,
    author = "Yoshimura, M.",
    title = "{Catastrophic particle production under periodic perturbation}",
    eprint = "hep-th/9506176",
    archivePrefix = "arXiv",
    reportNumber = "TU-484, TU-95-484",
    doi = "10.1143/PTP.94.873",
    journal = "Prog. Theor. Phys.",
    volume = "94",
    pages = "873--898",
    year = "1995"
}

@article{Felder:2006cc,
    author = "Felder, Gary N. and Kofman, Lev",
    title = "{Nonlinear inflaton fragmentation after preheating}",
    eprint = "hep-ph/0606256",
    archivePrefix = "arXiv",
    doi = "10.1103/PhysRevD.75.043518",
    journal = "Phys. Rev. D",
    volume = "75",
    pages = "043518",
    year = "2007"
}

@article{Frolov:2010sz,
    author = "Frolov, Andrei V.",
    title = "{Non-linear Dynamics and Primordial Curvature Perturbations from Preheating}",
    eprint = "1004.3559",
    archivePrefix = "arXiv",
    primaryClass = "gr-qc",
    reportNumber = "SCG-2010-04",
    doi = "10.1088/0264-9381/27/12/124006",
    journal = "Class. Quant. Grav.",
    volume = "27",
    pages = "124006",
    year = "2010"
}

@article{Greene:1997fu,
    author = "Greene, Patrick B. and Kofman, Lev and Linde, Andrei D. and Starobinsky, Alexei A.",
    title = "{Structure of resonance in preheating after inflation}",
    eprint = "hep-ph/9705347",
    archivePrefix = "arXiv",
    reportNumber = "SU-ITP-97-19, IFA-97-29",
    doi = "10.1103/PhysRevD.56.6175",
    journal = "Phys. Rev. D",
    volume = "56",
    pages = "6175--6192",
    year = "1997"
}

@article{Garcia-Bellido:2008ycs,
    author = "Garcia-Bellido, Juan and Figueroa, Daniel G. and Rubio, Javier",
    title = "{Preheating in the Standard Model with the Higgs-Inflaton coupled to gravity}",
    eprint = "0812.4624",
    archivePrefix = "arXiv",
    primaryClass = "hep-ph",
    reportNumber = "IFT-UAM-CSIC-08-93",
    doi = "10.1103/PhysRevD.79.063531",
    journal = "Phys. Rev. D",
    volume = "79",
    pages = "063531",
    year = "2009"
}

@article{Hertzberg:2014iza,
    author = "Hertzberg, Mark P. and Karouby, Johanna and Spitzer, William G. and Becerra, Juana C. and Li, Lanqing",
    title = "{Theory of self-resonance after inflation. I. Adiabatic and isocurvature Goldstone modes}",
    eprint = "1408.1396",
    archivePrefix = "arXiv",
    primaryClass = "hep-th",
    reportNumber = "MIT-CTP-4571",
    doi = "10.1103/PhysRevD.90.123528",
    journal = "Phys. Rev. D",
    volume = "90",
    pages = "123528",
    year = "2014"
}

@article{Kaiser:1997mp,
    author = "Kaiser, David I.",
    title = "{Preheating in an expanding universe: Analytic results for the massless case}",
    eprint = "hep-ph/9702244",
    archivePrefix = "arXiv",
    reportNumber = "HUTP-97-A005",
    doi = "10.1103/PhysRevD.56.706",
    journal = "Phys. Rev. D",
    volume = "56",
    pages = "706--716",
    year = "1997"
}

@article{Ellis:2025zrf,
    author = "Ellis, John and Garcia, Marcos A. G. and Olive, Keith A. and Verner, Sarunas",
    title = "{Constraints on Attractor Models of Inflation and Reheating from Planck, BICEP/Keck, ACT DR6, and SPT-3G Data}",
    eprint = "2510.18656",
    archivePrefix = "arXiv",
    primaryClass = "hep-ph",
    reportNumber = "UMN-TH-4512/25, FTPI-MINN-25/14, KCL-PH-TH/2025-42, CERN-TH-2025-199",
    month = "10",
    year = "2025"
}

@article{Garcia:2024rwg,
    author = "Garcia, Marcos A. G. and Pereyra-Flores, Aline",
    title = "{Impact of dark sector preheating on CMB observables}",
    eprint = "2403.04848",
    archivePrefix = "arXiv",
    primaryClass = "astro-ph.CO",
    doi = "10.1088/1475-7516/2024/08/043",
    journal = "JCAP",
    volume = "08",
    pages = "043",
    year = "2024"
}

@article{Garcia:2023dyf,
    author = "Garcia, Marcos A. G. and Gross, Mathieu and Mambrini, Yann and Olive, Keith A. and Pierre, Mathias and Yoon, Jong-Hyun",
    title = "{Effects of fragmentation on post-inflationary reheating}",
    eprint = "2308.16231",
    archivePrefix = "arXiv",
    primaryClass = "hep-ph",
    reportNumber = "UMN--TH--4223/23, FTPI--MINN--23/15, DESY-23-122",
    doi = "10.1088/1475-7516/2023/12/028",
    journal = "JCAP",
    volume = "12",
    pages = "028",
    year = "2023"
}

@article{Garcia:2023eol,
    author = "Garcia, Marcos A. G. and Pierre, Mathias",
    title = "{Reheating after inflaton fragmentation}",
    eprint = "2306.08038",
    archivePrefix = "arXiv",
    primaryClass = "hep-ph",
    reportNumber = "DESY-23-075",
    doi = "10.1088/1475-7516/2023/11/004",
    journal = "JCAP",
    volume = "11",
    pages = "004",
    year = "2023"
}

@article{Casagrande:2023fjk,
    author = "Casagrande, Gabriele and Dudas, Emilian and Peloso, Marco",
    title = "{On energy and particle production in cosmology: the particular case of the gravitino}",
    eprint = "2310.14964",
    archivePrefix = "arXiv",
    primaryClass = "hep-th",
    reportNumber = "CPHT-RR064.102023",
    doi = "10.1007/JHEP06(2024)003",
    journal = "JHEP",
    volume = "06",
    pages = "003",
    year = "2024"
}

@article{Mirzagholi:2019jeb,
    author = "Mirzagholi, Leila and Maleknejad, Azadeh and Lozanov, Kaloian D.",
    title = "{Production and backreaction of fermions from axion-$SU(2)$ gauge fields during inflation}",
    eprint = "1905.09258",
    archivePrefix = "arXiv",
    primaryClass = "hep-th",
    doi = "10.1103/PhysRevD.101.083528",
    journal = "Phys. Rev. D",
    volume = "101",
    number = "8",
    pages = "083528",
    year = "2020"
}

@article{Adshead:2022ecl,
    author = "Adshead, Peter and Liu, Aike and Lozanov, Kaloian D.",
    title = "{Production and backreaction of massive fermions during axion inflation with non-Abelian gauge fields}",
    eprint = "2203.09370",
    archivePrefix = "arXiv",
    primaryClass = "hep-ph",
    doi = "10.1088/1475-7516/2022/09/043",
    journal = "JCAP",
    volume = "09",
    pages = "043",
    year = "2022"
}

@article{Adshead:2018oaa,
    author = "Adshead, Peter and Pearce, Lauren and Peloso, Marco and Roberts, Michael A. and Sorbo, Lorenzo",
    title = "{Phenomenology of fermion production during axion inflation}",
    eprint = "1803.04501",
    archivePrefix = "arXiv",
    primaryClass = "astro-ph.CO",
    doi = "10.1088/1475-7516/2018/06/020",
    journal = "JCAP",
    volume = "06",
    pages = "020",
    year = "2018"
}

@article{Baacke:1998di,
    author = "Baacke, Jurgen and Heitmann, Katrin and Patzold, Carsten",
    title = "{Nonequilibrium dynamics of fermions in a spatially homogeneous scalar background field}",
    eprint = "hep-ph/9806205",
    archivePrefix = "arXiv",
    reportNumber = "DO-TH-98-10",
    doi = "10.1103/PhysRevD.58.125013",
    journal = "Phys. Rev. D",
    volume = "58",
    pages = "125013",
    year = "1998"
}

@article{Han:2025cwk,
    author = "Han, Jeonghak and Lee, Hyun Min and Song, Jun-Ho",
    title = "{Higgs pole inflation with loop corrections in light of ACT results}",
    eprint = "2506.21189",
    archivePrefix = "arXiv",
    primaryClass = "hep-ph",
    month = "6",
    year = "2025"
}

@article{Kazakov:2023tii,
    author = "Kazakov, D. I. and Iakhibbaev, R. M. and Tolkachev, D. M.",
    title = "{Leading all-loop quantum contribution to the effective potential in the inflationary cosmology}",
    eprint = "2308.03872",
    archivePrefix = "arXiv",
    primaryClass = "hep-th",
    doi = "10.1088/1475-7516/2023/09/049",
    journal = "JCAP",
    volume = "09",
    pages = "049",
    year = "2023"
}

\end{document}